 \documentclass{aa}  
\usepackage{natbib}
\usepackage{pdfpages}   
\usepackage[binary-units=true]{siunitx}
\bibpunct{(}{)}{;}{a}{}{,}
\renewcommand\deg{\ensuremath{}^\circ}

\usepackage{graphicx}
\usepackage{xcolor}
\usepackage{hyperref}
\usepackage{txfonts}

\usepackage{soul}
\newcommand{\hrieuv}{HRI\textsubscript{EUV}\xspace}
\newcommand{\hrilya}{HRI\textsubscript{Lya}\xspace}

\begin{document}

   \title{Stereoscopic Measurements of Coronal Doppler Velocities}

\author{O. Podladchikova\inst{\ref{i:pmod}}\fnmsep\thanks{Corresponding author: Olena Podladchikova \email{elena.podladchikova@pmodwrc.ch}}
        \and
        L. Harra\inst{\ref{i:pmod},\ref{i:eth}}
        \and
        K.Barczynski\inst{\ref{i:pmod},\ref{i:eth}}
         \and
        C.H. Mandrini\inst{\ref{i:uba}}
         \and
        F. Auch\`ere\inst{\ref{i:ias}}
        \and
        D. Berghmans \inst{\ref{i:rob}}
          \and 
        É. Buchlin\inst{\ref{i:ias}}  
           \and 
        L. Dolla\inst{\ref{i:rob}}
           \and   
        M. Mierla \inst{\ref{i:rob}, \ref{i:igar}}
          \and
        S. Parenti \inst{\ref{i:ias}}  
           \and 
        L. Rodriguez \inst{\ref{i:rob}}
    }
    \institute{
            Physikalisch-Meteorologisches Observatorium Davos, World Radiation Center, 7260, Davos Dorf, Switzerland\label{i:pmod}
            \and
            ETH-Z\"urich, Wolfgang-Pauli-Str. 27, 8093 Z\"urich, Switzerland\label{i:eth}
            \and
            Instituto de Astronomia y Fisica del Espacio, IAFE, UBA-CONICET, Ciudad Universitaria, CC. 67, 1428,  Buenos Aires, Argentina\label{i:uba}
             \and
            Université Paris-Saclay, CNRS, Institut d'Astrophysique Spatiale, 91405, Orsay, France\label{i:ias}
            \and
            Solar-Terrestrial Centre of Excellence -- SIDC, Royal Observatory of Belgium, Ringlaan -3- Av. Circulaire, 1180 Brussels, Belgium\label{i:rob}
            \and
            Institute of Geodynamics of the Romanian Academy, 19-21 Jean-Louis Calderon, 020032, Bucharest, Romania\label{i:igar}}
   
   \date{Received February 02, 2021; accepted July 30, 2021}

% \abstract{}{}{}{}
% 5 {} token are mandatory

  \abstract
   {The Solar Orbiter mission, with an orbit outside the Sun–Earth line and leaving the ecliptic plane, opens up opportunities for the combined analysis of measurements obtained by solar imagers and spectrometers. For the first time, different space spectrometers will be located at wide angles to each other, allowing three-dimensional (3D) spectroscopy of the solar atmosphere. }
{The aim of this work is to prepare the methodology to facilitate the reconstruction of 3D vector velocities from two stereoscopic LOS Doppler velocity measurements using the Spectral Imaging of the Coronal Environment (SPICE) on board the Solar Orbiter and the near-Earth spectrometers, while widely separated in space.}
{We develop the methodology using the libraries designed earlier for the STEREO mission but applied to spectroscopic data from the Hinode mission and the Solar Dynamics Observatory. We use well-known methods of static and dynamic solar rotation stereoscopy and the methods of EUV stereoscopic triangulation for optically-thin coronal EUV plasma emissions. We develop new algorithms using analytical geometry in space to determine the 3D velocity in coronal loops.}
{We demonstrate our approach with the reconstruction of 3D velocity vectors in plasma flows along "open" and "closed" magnetic loops. This technique will be applied to an actual situation of two spacecraft at different separations with spectrometers on board (SPICE versus the Interface Region Imaging Spectrograph (IRIS) and Hinode imaging spectrometer) during the Solar Orbiter nominal phase. We summarise how these observations can be coordinated.}
{}


   \maketitle
%
%----------------------------------------------------------
%%%%%%%%%%%%%%%%%%%%%%%%%%%%%%%%%%%%%%%%%%%%%%%%%%%%%%%%%%%%%%%
\section{Introduction}
\label{sec:intro}
%%%%%%%%%%%%%%%%%%%%%%%%%%%%%%%%%%%%%%%%%%%%%%%%%%%%%%%%%%%%%%%

%%%%%%%%%%%%%%%%%%%%%%%%%%%%%%%%%%%%%%%%%%%%%%%%%%%%%%%%%%%%%%%
The Solar Orbiter (SO) space mission \citep{SO}, launched in February 2020, provides both remote sensing and in situ measurements of the solar atmosphere and heliosphere. The main goal of this mission is to understand how the heliosphere is formed and sustained. The trajectory of Solar Orbiter  takes it out of the Earth’s orbit and into an orbit around the Sun, reaching  within  the orbit  of Mercury to 0.28 AU. A spacecraft whose orbit is away from the Sun–Earth line opens up huge possibilities for the representation of the three-dimensional (3D) imaging and spectroscopic data together with near-Earth imaging and spectroscopy. For the first time, spectrometers will be located at different angles from each other, allowing 3D spectroscopy of the solar atmosphere.

The presence of persistent  high-temperature, high-speed  upflows from the edges of active regions \citep{Harra2008} is a key discovery from Hinode \citep{EIS}. Measurements from the Extreme Ultraviolet (EUV) Imaging Spectrometer (EIS) indicate that  the upflows reach velocities of 50 km s$^{-1}$ with  spectral line asymmetries approaching 100 km s$^{-1}$  and more
 (see, {\it e.g.}, \citealt{Dolla2011}). It has been suggested that these upflows may lie on open magnetic field lines that connect to the heliosphere and may be a significant source of the low-speed  solar wind  \citep[][and references therein]{Harra2008,Brooks2011,Mandrini2014}. All of active regions observed by Hinode/EIS show upflows. Different  explanations have been given for the physical mechanism of the upflows, including waves, reconnection in the corona \citep{deb,Mandrini2015}, and reconnection in the chromosphere \citep{bart} driving energy upwards. The blue shifts in these regions are ubiquitous, and indicate the presence of upflows. Various studies have used modelling to determine whether the upflows that are seen at the edges of the active regions become plasma outflowing into the solar wind ({\it e.g.} \citealt{Boutry2012,edwards}).

Attempts have been made to understand how these flows vary with the location on the disc. Limb-to-limb studies of an active region were carried out by \citet{Demoulin2013} and \citet{Baker2017}. The highest plasma velocities in the three spectral lines that they explored have similar magnitudes, and their magnitudes increase with temperature. The authors concluded that their results are compatible with the active-region upflows originating from reconnection between active-region loops and neighbouring loops. However, having two spectroscopic views of an active region will enhance our understanding of flows.

First work on solar stereoscopy was carried out by assuming temporal stability of the features under study and allowing the Sun to rotate in order to obtain more than one viewpoint (e.g. \citealt{Berton1985,Koutchmy1992,AschwandenBastian1994,Aschwanden1995,Aschwanden1999,Feng2007a}). Such an approach is used also now \citep[see, e.g., ][]{Nistico2013}. The review by \citet{Aschwanden2011b} presents in detail the methods of static and dynamic solar rotation stereoscopy for coronal loops observed in optically-thin coronal plasma EUV emissions.

Two simultaneous viewpoints of the Sun were provided by the Solar Terrestrial Relations Observatory \citep[STEREO, ][]{Kaiser2008}. The two STEREO spacecraft orbit the Sun with increasing separation angles, providing stereoscopic images of the Sun’s atmosphere. The stereoscopic images obtained by the two nearly identical EUV broad-band imagers of the Sun Earth Connection Coronal and Heliospheric (SECCHI) suite \citep{Howard2008,wuelser04,Eyles2009} on board STEREO helped us to understand the 3D geometry of a rich variety  of optically  thin  solar structures \citep{Liewer2009, Patsourakos2009,Aschwanden2009, West2011, Delannee2014,Podladchikova2019, Mierla2008,Aschwanden2009,Mierla2009,Temmer2009,Mierla2010,Feng2009,dePatoul2013}.
 Methods of stereoscopic triangulation using a stereoscopic pair of EUV or white light coronal images have been developed. A lot of progress has been made to improve magnetic field models based on stereoscopic information, and to beat down the discrepancy between theoretical magnetic field models and observed stereoscopically triangulated loop 3D coordinates \citep{Aschwanden2011b} .

Several studies  have been carried out in parallel to calculate the true location of coronal structures in 3D space  \citep{Pizzo2004,Inhester2006,Feng2007b,Aschwanden2008,HowardTappin2008}. These are based on the direct geometric triangulation  using a series  of line-of-sight (LOS) measurements taken from different spacecraft views towards the apparent edges of the structures. The true 3D coordinates of the structures are calculated from the intersections of these LOSs. The theoretical background for solar stereoscopy based on the direct triangulation of solar structures is given by \citet{Inhester2006}. The direct triangulation technique can be applied after prior identification and matching of the targeted structures in two images (see the review by \citealt{Wiegelmann2009}). The alternative method of magnetic stereoscopy has been proposed by \cite{Wiegelmann2002}. The success of magnetic stereoscopy depends on the quality of theoretical magnetic field models. A critical assessment of non-linear force-free field (NLFFF) models identified a substantial mismatch between theoretical magnetic field models extrapolated from photospheric magnetograms and stereoscopically triangulated loops, in the order of a 3D misalignment angle of $\alpha_{mis}\approx 20^\circ$ to  $40^\circ$ \citep{DeRosa2009,Sandman2009,Aschwanden2011b}. \citet{Rodriguez2009} highlighted that every pixel in an image is a result of the LOS integration of the emission of the optically thin coronal plasma. This problem of LOS integration persists in the analysis of the EUV data, further complicating the matching of points between the two images.

The triangulation method has been applied to  detailed 3D reconstructions of coronal loops in active regions obtained from a stereoscopic pair of images  \citep{Inhester2006,Aschwanden2008,Aschwanden2011a,Aschwanden2012a,Nistico2013,Aschwanden2015,Chifu2017}. An  example of such work \citep{Rodriguez2009} reveals that loops that appear to be co-spatial in 171~\AA\ and 195~\AA\ images have in fact different heights and occupy different volumes. These results are key to understanding coronal heating.

Here, we develop a 3D EUV spectroscopy methodology for active regions that is used to build velocity vectors from a pair of EUV images and from Doppler shift maps taken from different perspectives. We do this using the STEREO triangulation technique of EUV images of active regions and we develop novel methods of analytical geometry in space to determine 3D velocities in coronal loops. An advantage of having two different views of an active region is that one can determine the projection angles of the loops, and hence direction of plasma flows along them. We use Hinode/EIS data acquired at different times in order to replicate the situation of two spacecraft with spectrometers on board, and we choose an active region that did not show significant changes in time (Section~\ref{STEP1}). In addition, we describe methods with which to reconstruct the velocity vectors; these will be released as a suite of 3D spectroscopy algorithms (Software package DOVES –– DOppler VElocities Stereoscopically) for velocity vector reconstruction. We describe the spatio-temporal co-alignment  between EUV broad-band and spectroscopic images (Section~\ref{STEP2}), the reconstruction of the 3D geometry of coronal loops (Section ~\ref{STEP3}), and deprojection algorithms of the measured LOS Doppler shifts into velocity vectors of plasma flows onto straight and curved coronal loop structures (Section~\ref{STEP4}). We also perform magnetic field modelling of the active region to investigate the possibility of  using it to derive the 3D loop geometry (Section~\ref{B_model}).

A description of the instruments that could be  used for 3D spectroscopy is given in Section 6. We provide recommendations on the optimum possible spacecraft configuration and spatial resolution of the instruments for 3D spectroscopy within the framework of the Solar Orbiter  mission. The proposed 3D spectroscopy method is of particular interest when aiming to understand how the corona is heated and how the solar wind is formed.

%%%%%%%%%%%%%%%%%%%%%%%%%%%%%%%%%%%%%%%%%%%%%%%%%%%%%%%%%%%%%%%%%%%

\section{Data preparation for 3D spectroscopy}
\label{STEP1}

\subsection{Instrumentation}
Initial data for 3D velocity reconstruction includes a stereoscopic pair of EUV images, and simultaneous Doppler shifts measurements from two angularly separated spectrometers. Now, to simulate the stereo view we get usable simultaneous data sets with SPICE, we use solar rotation to produce two different viewpoints, as it was done for STEREO sofware preparation. These viewpoints are imitated now by imaging data obtained by the Atmospheric Imaging Assembly (AIA) on board the Solar Dynamic Observatory (SDO, \citealt{SDO}) and spectroscopic data from Hinode/EIS. By applying direct triangulation methods to SDO/AIA intensity images, we restore the 3D coordinates of the observed structures. Then, the triangulation of the Doppler shifts observed by Hinode/EIS allows us to restore the vector velocities in these structures.

Hinode is a Japanese mission launched in 2006 by the Institute of Space and Astronautical Science (ISAS) of the  Japan Aerospace Exploration Agency (JAXA) in collaboration  with the National Astronomical Observatory of Japan   \citep{Kosugi2007}. Hinode/EIS has two wavelength bands, 170–211~\AA\ and 246–292~\AA, that include spectral lines formed over a wide range of temperatures, from chromospheric to flare temperatures. Hinode/EIS has an effective spatial resolution of about 3–4 arcsec. The high spectral resolution (0.06~\AA) allows the determination of Doppler velocity maps with an accuracy of  3 km s$^{-1}$. The Hinode/EIS campaign used
in this work has a slit size of 1 arcsec, a field of view (FOV) of 467 arcsec $\times$ 511 arcsec and a raster duration of 67 min.

The SDO/AIA, built  by Lockheed Martin Solar and Astrophysics Laboratory, provides continuous broad-band full-disc images from the chromosphere to the solar corona in seven EUV wavelengths covering the temperature range $2 \times 10^4-2 \times 10^7$~K, with a cadence of
12 s and a spatial resolution of 1.2 arcsec (corresponding to two pixels). SDO is in a circular geosynchronous orbit at an altitude of 35\,800 km.

Active region NOAA AR 2678 was observed simultaneously by SDO/AIA  and Hinode EIS on 19 and 22 November 2017. The separation of the two observing positions due to solar rotation is 37$\deg$, which is within  the angular 
%distance 
range $9\deg-120\deg$ in which stereoscopic vision and 3D spectroscopy becomes applicable. We discuss this aspect in Section~\ref{STEP2}.

\subsection{Observations of NOAA AR 2678}
	%%%%%%%%%%%%%%%%%%%%%%%%%%%%%%%%%%%%%%%%%%%%%%%%%%%%%%%%%%%%%%%%%%%
	\begin{figure}[!hbt]
			\centerline{
			\includegraphics[width=1.0 \linewidth,clip=]{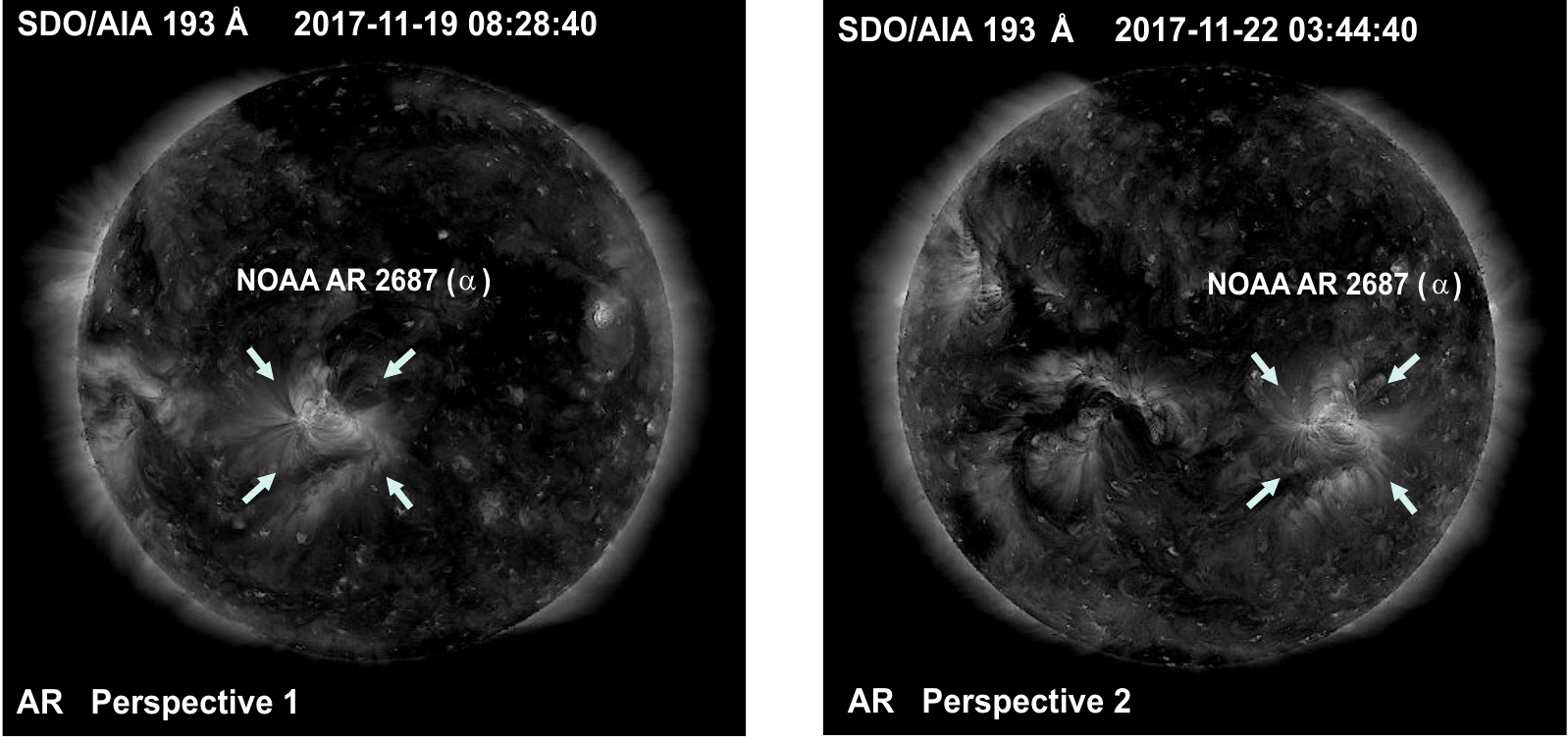}
		}	
		\caption{Whole Sun images obtained with  SDO/AIA at 193~\AA\ on 19 November 2019 (position  1, left) and 22 November 2017 (position 2, right). The active region (pointed with
		arrows) has a simple magnetic configuration ($\alpha$ according to the Zurich classification). These two views correspond to a separation of $37\deg$  between position 1 and position 2, allowing the use of stereoscopy for the 3D reconstruction of the active region. 
 We illustrate the active region dynamic evolution with four animations:
193~\AA\ from 19 to 22 November \href{https://drive.google.com/file/d/1DtK4syJ1oF88HB9eQJKgzIwVx8lb_XIe/view?usp=sharing}{ ({\color{blue}Movie 1})}, 
stabilized 193~\AA\ from 19 to 22 November \href{https://drive.google.com/file/d/1poUe4ymyPZHPEIcf67C_xCMQgKlkm4og/view?usp=sharing}{ ({\color{blue} Movie 2})}, 
2D SDO/HMI photospheric magnetic field with overlaid 193~\AA \href{https://drive.google.com/file/d/1Ev8sTHAW_N4w13Iv-187tZDR7QGLYxK2/view?usp=sharing}{ ({\color{blue} Movie 3})}, 
171~\AA\ \href{https://drive.google.com/file/d/1u-o6O9gt8xHpO8CbaQMA5T8L48TZTLwC/view?usp=sharing}{ ({\color{blue} Movie 4})}. 
Animations demonstrate that some of the observed simple shape loops can be traced throughout the entire animation. 
}
		\label{Fig1}
	\end{figure}
	%%%%%%%%%%%%%%%%%%%%%%%%%%%%%%%%%%%%%%%%%%%%%%%%%%%%%%%%%%%%%%%%
%

On each viewpoint, three simultaneous  images were obtained: (1) \ion{Fe}{xii}  intensity from SDO/AIA, (2) \ion{Fe}{xii} intensity  from Hinode EIS and (3) \ion{Fe}{xii} Doppler velocity map. 
%Both SDO/AIA and Hinode EIS data are used for 3D spectroscopy; 
The SDO/AIA images are used to establish the 3D coordinates of coronal points and the Hinode EIS images for velocity vector construction in 3D space. Figure~\ref{Fig1} shows whole Sun images obtained with SDO/AIA at 193~\AA. 
Because the active region evolves slowly, its apparent morphological changes are due to the different viewing angles.

\section{Spatio-temporal data co-alignment for 3D spectroscopy}
\label{STEP2}
\subsection{Hinode/EIS data}

We used the eis\_prep.pro routine in SolarSoftware to calibrate the Hinode/EIS data. The dark current and cosmic rays were removed, and the hot pixels were corrected. The intensity DN (digital number) values were calibrated to the spectral radiance in units of erg $\cdot$ (cm$^2$~$\cdot$~s~$\cdot$~ sr~$\cdot$~\textrm{\AA})$^{-1}$. We focused on the strongest emission line observed by Hinode EIS, \ion{Fe}{xii} at 195.12~\AA\ with formation temperatures $\log T=6.1$. For each spectrum, we fitted a single Gaussian function with the eis\_auto\_fit routine and obtained the line peak intensity, Doppler velocity, and line width using eis\_get\_fitdata.pro.

\subsection{SDO/AIA data processing}
 We have chosen the following conditions for the SDO/AIA 193~\AA\ data, the observation time corresponds to  that  of the whole Hinode/EIS raster scan. The FOV  is slightly larger than the region observed by Hinode/EIS and it is tracked with solar rotation at the Carrington rate. Images were transformed to a grid with a resolution of 0.6 arc seconds per pixel. The pre-processed data were aligned with the solar north and divided by the exposure time. The strongly saturated frames were removed. We  selected SDO/AIA 193~\AA\  images with  more than 2~s of exposure time to obtain suitable images for long-lived coronal structure analysis. We processed the images with an image stacking technique to increase photon-to-noise statistics and then with a wavelet high-pass filter to enhance coronal structures \citep{Stenborg2008}.

\subsubsection{Hinode/EIS and SDO/AIA  data co-alignment}

Dynamic structures on the solar surface evolve with a significantly shorter time than the duration of the Hinode/EIS raster observation of 67 min. We  created a pseudo SDO/AIA  raster  analogous to the Hinode/EIS raster to compensate  for possible dynamic changes. 
%during the Hinode EIS raster. 
To this end, we extracted the SDO/AIA data closest in time to each position of the Hinode/EIS slit. %\cmc{Difficult to understand}.
These data were merged into a single SDO/AIA  map (hereafter referred to as “artificial SDO/AIA raster map”), corresponding to Hinode/EIS raster map using the cross-correlation method described herein. The offset between the artificial  SDO/AIA  raster map and the Hinode/EIS raster map was applied together with a pointing correction. Based on the new Hinode/EIS pointing details, the new artificial SDO/AIA raster map was created and aligned to the new Hinode/EIS map. In our analysis, we set the expected accuracy as half the size of the Hinode/EIS pixel.
%
	%%%%%%%%%%%%%%%%%%%%%%%%%%%%%%%%%%%%%%%%%%%%%%%%%%%%%%%%%%%%%%%%%%%
	\begin{figure}[!hbt]
			\centerline{
			\includegraphics[width=1.0 \linewidth,clip=]{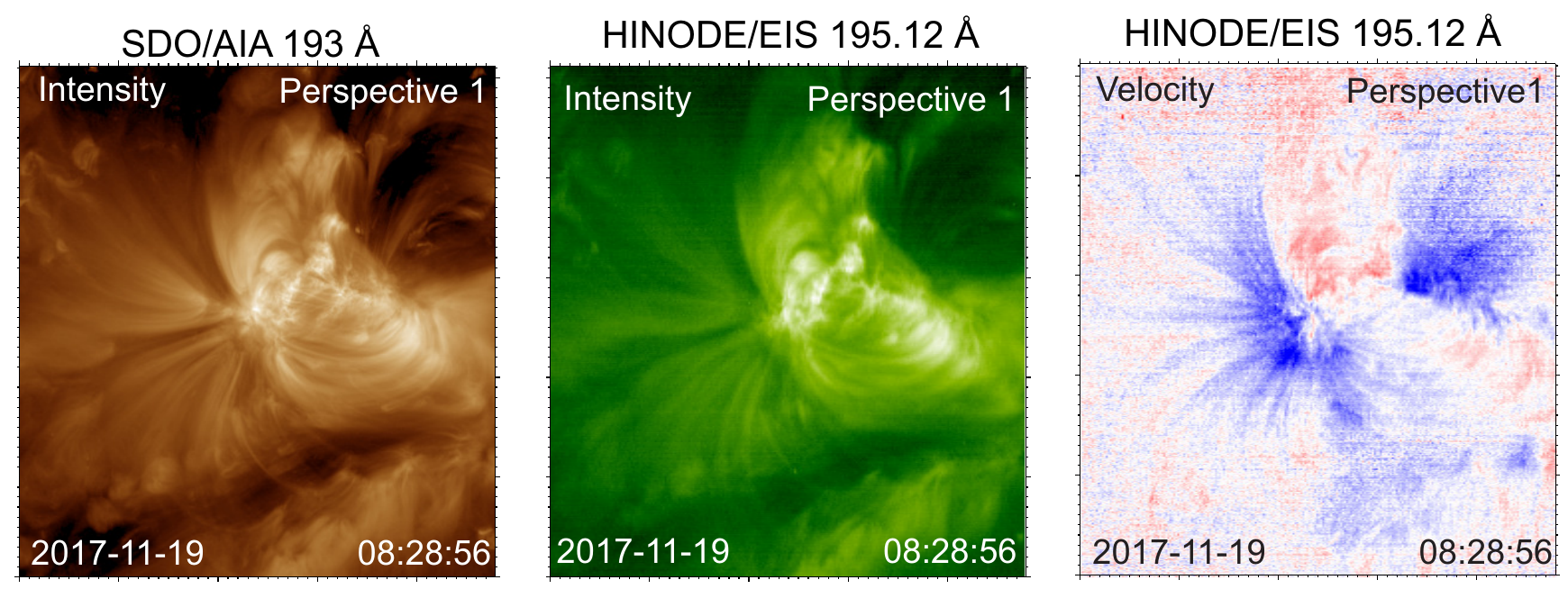}
		}	
		\caption{Hinode/EIS intensity (middle) and Doppler velocity (right) raster maps with  corrected pointing  information, and the spatio-temporally  corresponding artificial SDO/AIA  raster map (left) obtained by the instruments in position 1.}
		\label{Fig2}
	\end{figure}
	%%%%%%%%%%%%%%%%%%%%%%%%%%%%%%%%%%%%%%%%%%%%%%%%%%%%%%%%%%%%%%%%

Figure~\ref{Fig2} shows the Hinode/EIS intensity and Doppler velocity raster maps and the corresponding artificial SDO/AIA raster map. A similar procedure was presented in \citet{Barczynski2018} to co-align SDO/AIA  and the Interface Region Imaging Spectrometer (IRIS) raster data. Here, 3D point triangulation is performed with SDO/AIA data, and each 3D point is assigned to a pair of Doppler shifts via spatio-temporal SDO/AIA -- Hinode/EIS co-alignment.

\section{3D triangulation of NOAA AR  2678}
\label{STEP3}

\subsection{Triangulation through epipolar geometry}

\begin{figure}[!hbt]
			\centerline{
			\includegraphics[width= 1.0 \linewidth,clip=]{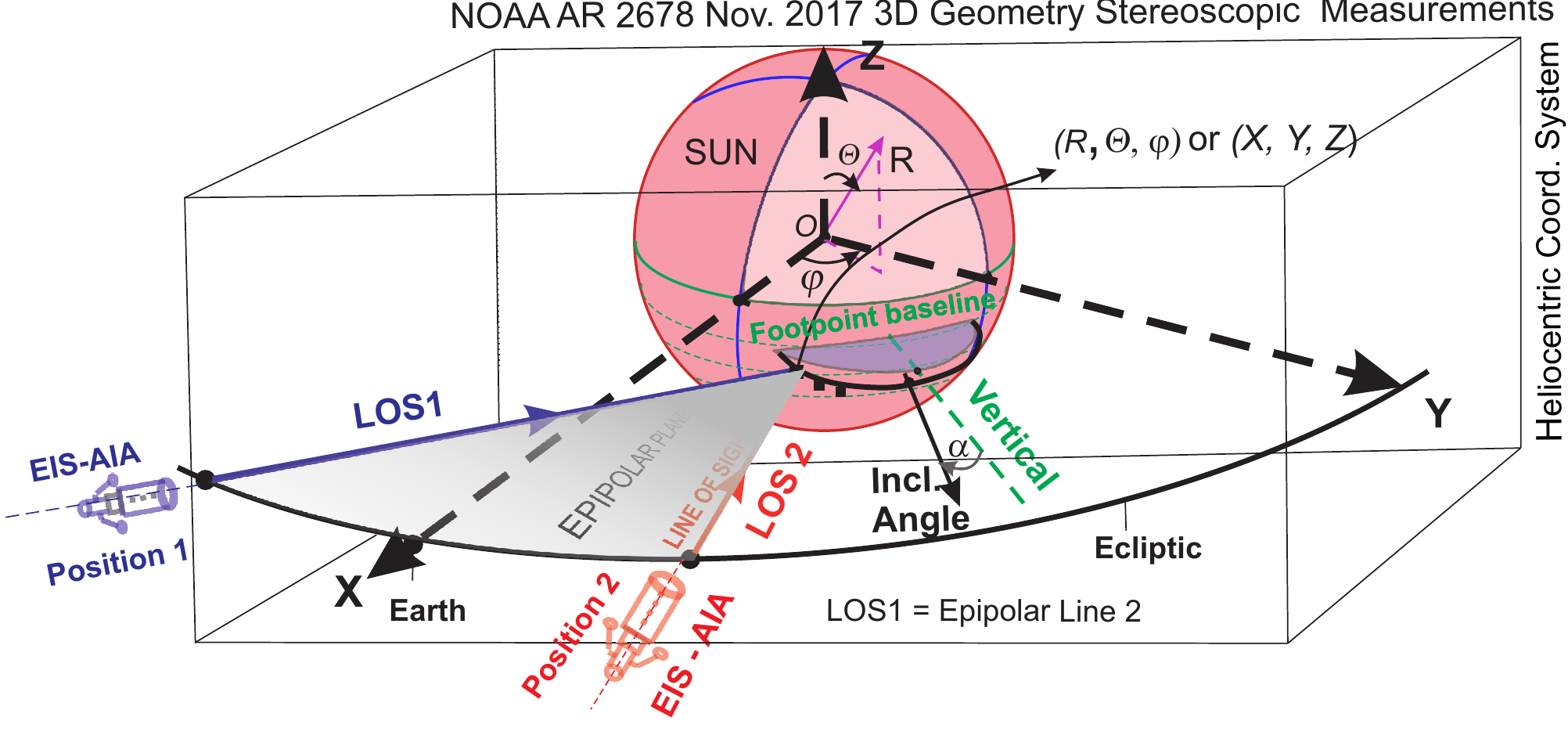}
		}
		\caption{
		Stereoscopic observation of NOAA AR 2678 with  two spacecraft. The positions of the two spacecraft and any point  in  the solar corona to  be triangulated define a plane called the epipolar plane \citep[see, e.g.,][]{Inhester2006}. Because the two LOSs  (LOS1 and LOS2) must lie in the same epipolar plane, their  intersection in this plane is unambiguously defined by the 3D point.  For convenience, the 3D coordinates of the points are reconstructed in HEEQ heliocentric coordinates (axis X is directed towards  the Earth).  Calculation of the 3D-point coordinates is the first  step in the velocity vector construction.
}
		\label{Fig33}
	\end{figure}

The 3D coordinates are calculated as (Earth-based) Stonyhurst heliographic longitude and latitude, along with the radial distance in solar radii. Then, the coordinates are converted into the Heliocentric Earth Equatorial (HEEQ) coordinates for a Cartesian representation of the data. A two-dimensional (2D) solar image taken by a spacecraft is usually identified by the $(i, j)$ coordinates in the image coordinate system or by  the latitude  and longitude in the heliographic (or helioprojective Cartesian) coordinate system, when $(i, j)$ are projected on the solar surface \citep{Thompson2006}. The multi-point heliospheric observation in the corona requires a complete 3D heliocentric coordinate system, as explained in Section~\ref{3Ddtri}.  Figure~\ref{Fig33} shows the HEEQ coordinate system with the origin in the centre of the Sun and the object position in the heliosphere described by the Cartesian coordinates $X, Y, Z $. The Stonyhurst coordinate system describes the 3D position of a feature with the spherical coordinates $ R, \Theta, \varphi$.

	%%%%%%%%%%%%%%%%%%%%%%%%%%%%%%%%%%%%%%%%%%%%%%%%%%%%%%%%%%%%%%%%%%%
	%%%%%%%%%%%%%%%%%%%%%%%%%%%%%%%%%%%%%%%%%%%%%%%%%%%%%%%%%%%%%%%%

As seen from Figure~\ref{Fig33}, two observing spacecraft positions together with a targeted point in the solar corona define a plane called the epipolar plane. All targeted points have planes in common that contain the two spacecraft positions. Given that every epipolar plane is seen head-on from both spacecraft, it is reduced to a line in the respective image projections; this line is called an epipolar line.

Any targeted corona point  identified to be situated on a certain epipolar line in one image must lie on the  same epipolar line in the other image. The epipolar lines therefore provide a natural coordinate system for stereoscopic reconstructions.
Consequently, finding a correspondence between pixels in the images taken by two EUV imagers separated in space is reduced to establishing a correspondence between pixels along the same epipolar lines in the two images.

Once the correspondence between the pixels is found, the 3D reconstruction is performed by calculating the LOSs that belong to the respective pixels in the image and back-tracking them into 3D space. Because the LOSs must lie in the same epipolar plane, their intersection in this plane is defined unambiguously. This procedure is often called “tie-pointing” (see,  e.g., \citealt{Inhester2006,Mierla2009,Liewer2009,Aschwanden2011b}). The geometrical reconstruction errors are related to both the separation angle  $\gamma$ between the two spacecraft and the spatial resolution  $ds$ of an image \citep{Inhester2006,Aschwanden2015}; see Section~\ref{geom_err} for more information on reconstruction errors in the context of the Solar Orbiter mission.

\subsubsection{3D-triangulation software}
\label{3Ddtri}
\overfullrule=5pt

In this study, we use the HEEQ system. The origin of the HEEQ system is the intersection of the solar equator with the central meridian as seen from Earth, and the solar feature location is given either in Cartesian coordinates X, Y, Z (HEEQ coordinates) or in spherical coordinates (Stonyhurst heliographic coordinates: latitude $\Theta$, longitude $\varphi$  and heliospheric radius $R$)  \citep{Thompson2006} (see Figure~\ref{Fig33}). The locations of the features were processed  with  the World  Coordinate System routine scc\_measure.pro of the Solar Software  \citep{Thompson2006}. The structures were traced in stereoscopic pairs of images using projections along the epipolar lines. The routine uses a combination of the information in the header, such as the pixel-to-degree conversion,  and the SPICE (Spacecraft Planet Instrument C-matrix Events orientation) database of orbital  kernels by calling the routine convert\_sunspice\_coord.pro, which is based on SPICE orbital ephemeris kernels containing the spacecraft location and pointing information. The coordinate system of each spacecraft image plane can be related to a heliocentric coordinate system, and stereoscopic analysis is performed. The output of scc\_measure.pro is given in the Stonyhurst heliographic coordinates.

\subsubsection{3D triangulation of AR 2678 points}

\begin{figure}[!hbt]
			\centerline{
			\includegraphics[width=1.0 \linewidth,clip=]{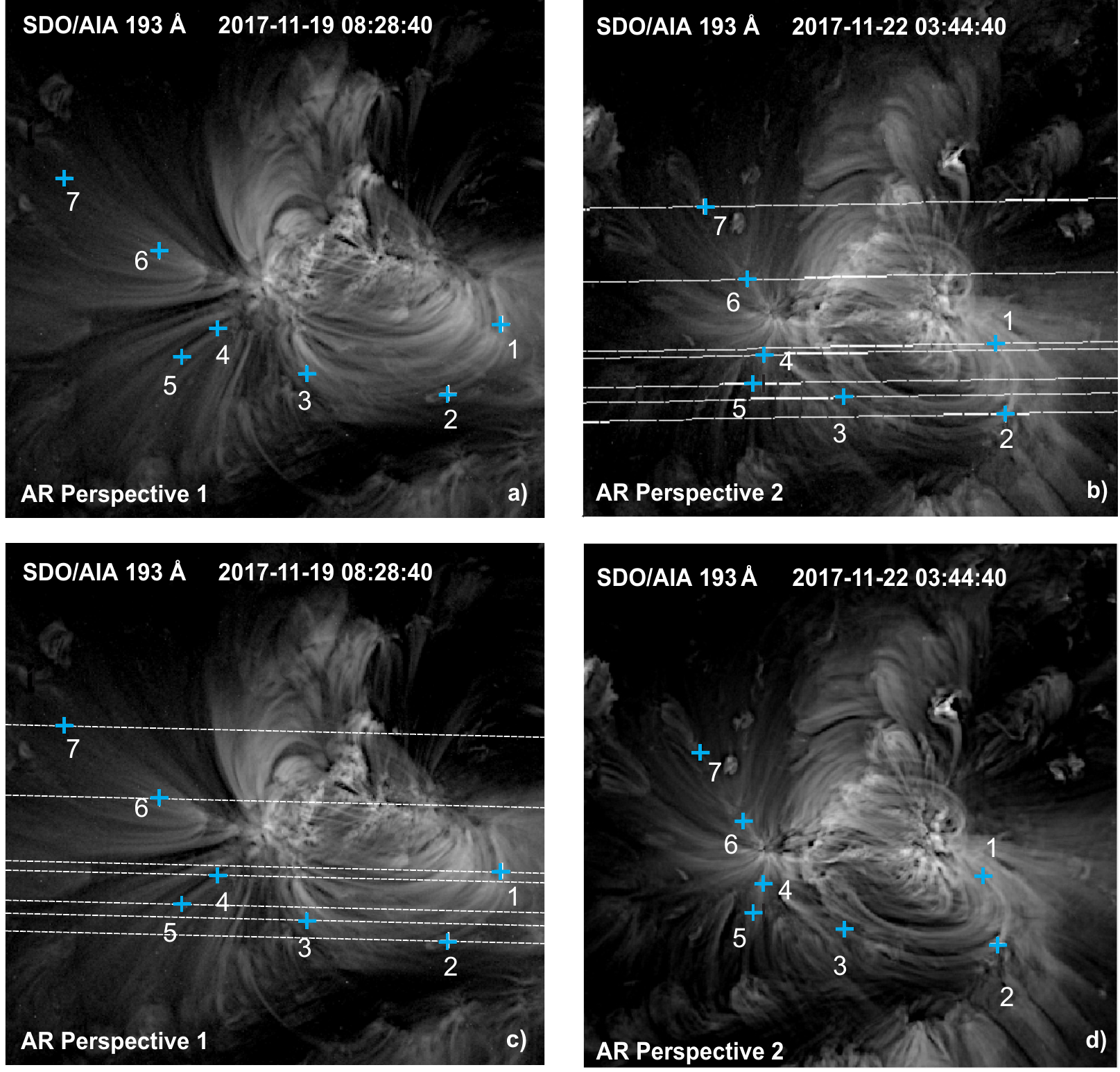}
		}	
		\caption{(a) Points 1  to 7  on Image 1 selected for triangulation. (b) Points 1 to 7  on Image 2 are found on the corresponding epipolar lines (dashed lines) at the intersection with  the observed structures. Points 1,  2 and  3  belong to closed loops (we call as such those for which we can identify both of their bases in AIA EUV images); whereas points 4, 5, 6, and 7  belong to open loops (i.e. those for which we can identify only one of their bases in AIA EUV images). To validate the tie-pointing procedure, we performed the reverse process: first starting from the obtained points on Image 2 (panel d) and then selecting their stereoscopic pairs on the corresponding epipolar lines on Image 1 (panel c), obtaining the same points.
        }
		\label{Fig4}
	\end{figure}

Direct measurement of the 3D coordinates of a point located high in the solar corona is not possible. However, if the two LOSs traced back from the LOS projections on the 2D image intersect higher in the corona at the observed point, then the 3D coordinates  of the point  can be evaluated. The procedure is performed as follows:
\begin{itemize}
\item Point selection on a 2D plane of sky (Image 1); see Figure~\ref{Fig4}a.
\item Transition to 3D coordinates.
\item LOS drawing from satellite 1 to the observed point (LOS 1 or
epipolar lines in stereoscopy). The LOSs are not visible on the 2D Image
1 (Figure~\ref{Fig4}a), because they are projected to one point.
\item Image 2 (Figure~\ref{Fig4}b) shows the same area in the solar corona from Perspective
2. LOS1 is visible in Image 2 (Figure~\ref{Fig4}b) due to observation from a  different point.
\item  
The intercept of LOS 1 with coronal loops, the structure of the observation, determines the projection of the 3D point on Image 2 (crosses in Figure~\ref{Fig4}b). Thus, the problem converges, as long as we deal with two LOS projections of the same point in 3D space.
\end{itemize}

To check the obtained coordinates the procedure can be then performed in reverse order (Figure~\ref{Fig4}c,d). In Figure~\ref{Fig4}d we start from the obtained points on Image 2 and describe how we arrive at the initially  selected points on Image 1 (Figure~\ref{Fig4}c) using the LOS projection technique. We repeat the procedure for points 1-2-3 located on the same loop (same for points 4-5 and 6-7) selected on the SDO / AIA images, and express their true spatial position in a 3D heliocentric coordinate system.
% if we fail we restart from the beginning.

	%%%%%%%%%%%%%%%%%%%%%%%%%%%%%%%%%%%%%%%%%%%%%%%%%%%%%%%%%%%%%%%%%%%
	%%%%%%%%%%%%%%%%%%%%%%%%%%%%%%%%%%%%%%%%%%%%%%%%%%%%%%%%%%%%%%%%

\subsubsection{3D loop parameterisation}

Here we study the loops, which are shown in Figure ~ \ref{Fig4} as straight, curved or semicircular, which is typical for loops that are inclined towards the surface near the center of the disc \citep{Reale2010}. \citet{Aschwanden2011b} presents the technique of semi-circular loop point positions and loop plane inclination definitions. The plane in the strict geometric sense has zero thickness; nevertheless, the terms “loop plane” and “loop plane inclination” are the established and widely used terms in the analysis of the parameters of coronal loops (see, e.g., Section "Stereoscopic Fitting of Circular Loop Geometry” in \citet{Aschwanden2009d}, or \citet{Nistico2013} and \citet{Aschwanden2011a}), even though the real loops are thick \citep{Klimchuk1992}.

From the 3D triangulation, we obtain several points on the same loop. We model the loop by a circle, as explained in \citet{Rodriguez2009} and \citet{Aschwanden2011b}, defining the curvature radius, and we use three loop points triangulated as described above to fit  the loop with the model. The loop can be defined in 3D space by a set of three points with coordinates $(X, Y, Z)$ in the HEEQ coordinate system and the two sets of 2D coordinates of the solar images $(i_{Los1},j_{Los1})$ and $(i_{Los2},j_{Los2}) $; however, it is possible to use more points to fit the circle. We save the projected  coordinates of the 3D points $(i_{Los},j_{Los})$ on the images obtained from the two perspectives for the subsequent operations with Doppler velocity measurements. More details on loop parameterisation are given in \citet{Aschwanden2015}, \citet{Aschwanden2011b} and \citet{Nistico2013}.

\begin{figure}[!hbt]
		\centerline{
			\includegraphics[width=0.8 \linewidth,clip=]{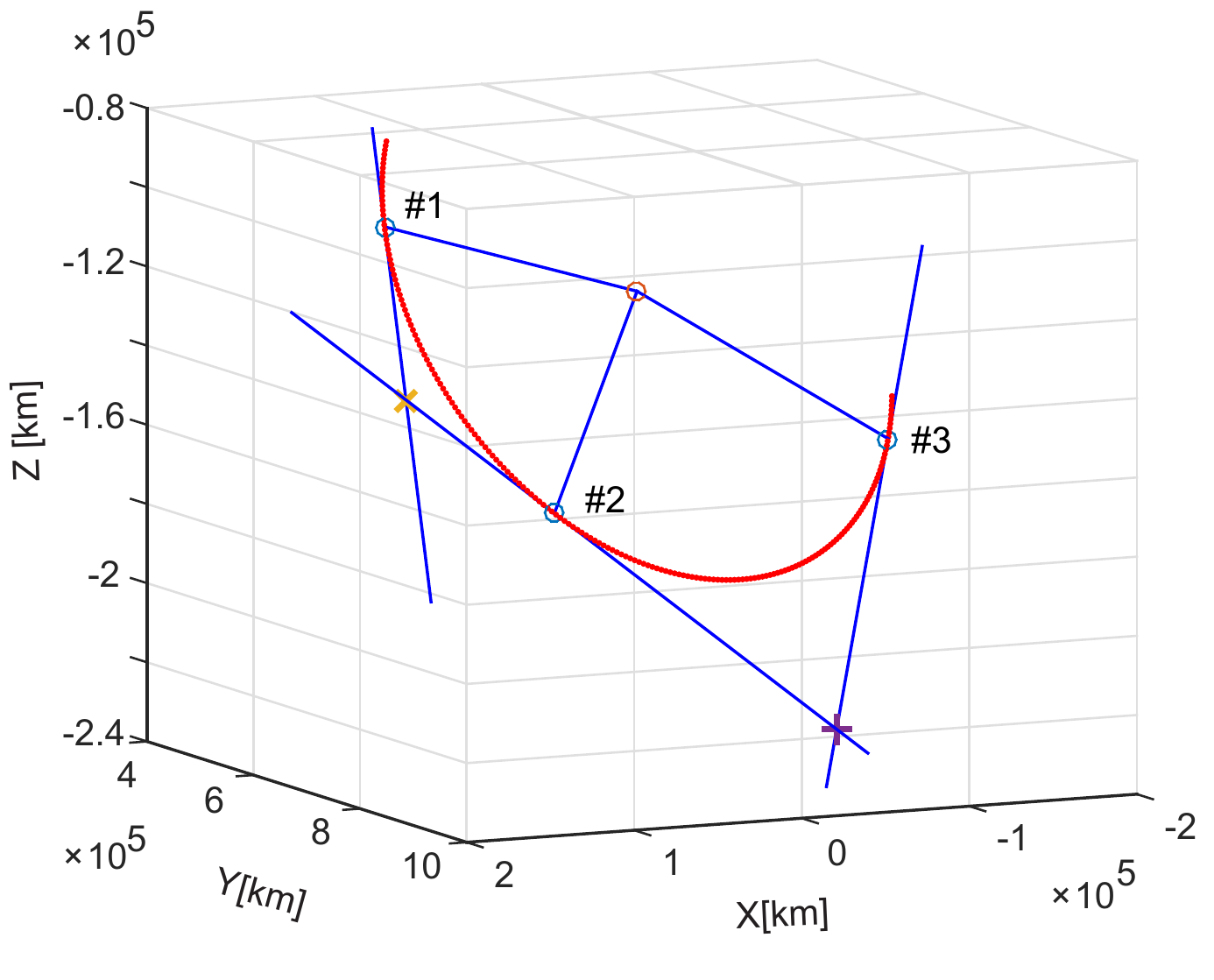}
		}	
		\caption{Semi-circular curved coronal loop under study observed stereoscopically in NOAA AR 2678 in November 2017. The loop is shown in the HEEQ coordinate system. Blue lines outside the loop show tangents to the loop plotted at the selected points.}
		\label{Fig5}
	\end{figure}

When working with optically thin objects in the solar corona, the most critical task is to ensure that two LOSs truly intersect on the object under study in 3D EUV images of the corona. Figure~\ref{Fig5} shows the coronal loop under study, presented  in the 3D HEEQ coordinate system and reconstructed  from points 1, 2 and 3.  Blue lines outside the loop show tangents to the loop plotted at the observed points.

%%%%%%%%%%%%%%%%%%%%%%%%%%%
	%%%%%%%%%%%%%%%%%%%%%%%%%%%%%%%%%%%%%%%%

\subsubsection{NOAA AR 2678 3D loops}
%%%%%%%%%%%%%%%%% TABLE 1 
	\begin{table*}[ht!]
	\centering
		\caption{
		Latitude, longitude,  and height above the photosphere of the seven selected points in Figure 4. The 1\textsuperscript{st} column indicates the point number. The 2\textsuperscript{nd}, 3\textsuperscript{rd}, and 4\textsuperscript{th} columns show the latitude and longitude from Earth,  Position  1 and Position 2 viewpoints, respectively. The 5\textsuperscript{th} column lists the height of the points above the photosphere, and the 6\textsuperscript{th} column shows the inclination angles $\alpha$ between the open coronal loop, or closed coronal loop plane, and the vertical  to the solar surface. See text for the meaning of closed and open loops.
}
		\begin{tabular}  {|p{1.8cm} l l l l p{1.8cm}|}
 %{| p{1.3cm} p{2.3cm} p{2.3cm} p{2.3cm} p{1.0cm} p{0.7cm}|}
	      \hline			
	        & Earth  &  Position 1 & Position 2 &  Height above & Inclination $\alpha$ \\
	        & (Lat./Long.) &  (Lat./Long.) & (Lat./Long.) & Photosphere & Degrees  \\
	       \hline 
			   {\bf Closed loop}  &  &  &  &  &    \\

              Point 1  &  $-10.7\deg/-3.04\deg$  &  $-13.2\deg/-3.13\deg$ & $-12.7\deg/35.16\deg$ &  $16.0$ Mm  &   \\

              Point 2  & $-12.9\deg/~~~5.2\deg$  &  $-16.6\deg/-5.9\deg$ & $-16.2\deg/36.40\deg$ &  $76.1$ Mm & {$15.5\deg$}  \\

              Point 3  & $-12.9\deg/-12.2\deg$   &    $-15.6\deg/-12.9\deg$ &   $-15.3\deg/26.2\deg $   & $29.6$ Mm &     \\
				
            {\bf Open loop} &  &   &   &  &   \\

            Point 4 & $-10.9\deg/-16.5\deg$ & $-13.4\deg/-17.24\deg$ & $-13.2\deg/21.3\deg$ & $22.3$ Mm   \\

             Point 5 &  $-11.7\deg/-17.5\deg$ & $-14.8\deg/-19.2\deg$ & $-14.6\deg/21.4\deg$ & $55.1$ Mm &
             $54.1\deg$  \\
	
           {\bf Open loop} & & & & &  \\

          Point 6 & $-6.9\deg/-18.5\deg$ & $-9.6\deg/-19.9\deg$ & $-9.3\deg/20.02\deg$ & $48.8$ Mm &       \\

           Point 7  & $-3.2\deg/-21.7\deg$  & $-5.90\deg/-24.7\deg$ & $-5.7\deg/17.4\deg$ & $89.1$ Mm & $55.5\deg$\\
            \hline
          \end{tabular}
          \label{Tloops}
\end{table*}
		%%%%%%%%%%%%%%%%% TABLE 1 <

Table~\ref{Tloops} summarises the results  of the 3D reconstructed points in the Stonyhurst coordinate system from the Earth point of view, as well as from the Position 1 and Position 2 perspectives. The first entry in the table lists the results for points selected on a closed loop within  the active  region core, where the dominant Doppler velocity indicates a downflow, and the bottom part of the table shows results for points on open loops at the edges of the active region, where the Doppler velocities show blue-shifted plasma (upflow). Notice that we use the terms closed loops for those for which we can identify their two bases in AIA EUV images, and open loops for those for which we can identify only one of their bases in AIA EUV images. 
%\cmc{This is a repetition of the text in caption to Figure 4, but I do not see how to avoid the repetition.} 

%%%%%%%%%%%%%%%%%%%%%%%%%%%%%%%%%%%%%%%%%%%	
\section{3D spectroscopy methodology}
\label{STEP4}
\subsection{Vector velocity measurements}
%
	%%%%%%%%%%%%%%%%%%%%%%%%%%%%%%%%%%%%%%%%%%%%%%%%%%%%%%%%%%%%%%%%%%%
	\begin{figure}[ht!]
			\centerline{
			\includegraphics[width=1.0 \linewidth,clip=]{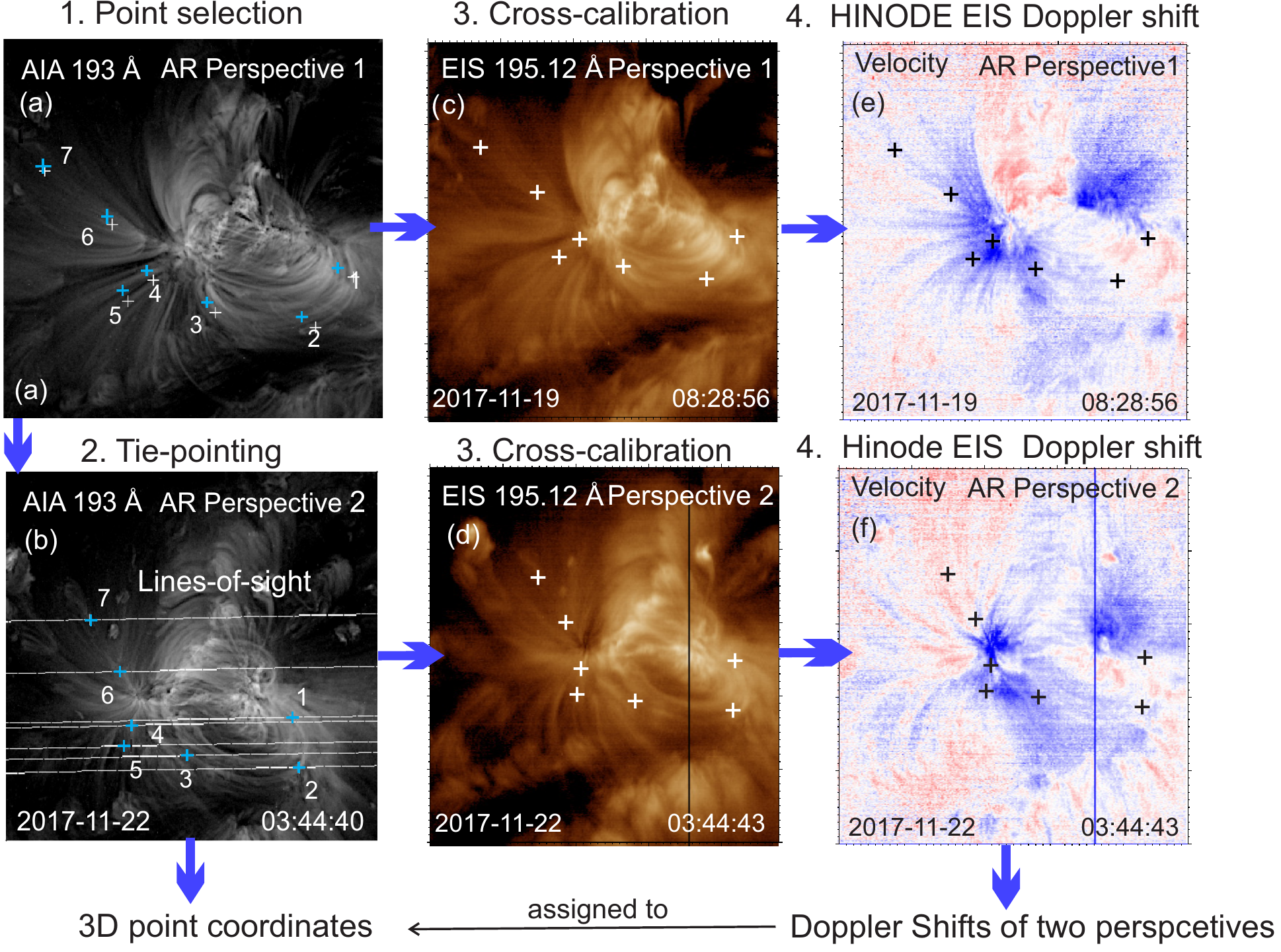}
		}	
		\caption{
		Stereoscopic view of NOAA  AR 2678, recorded from different perspectives in November 2017. (a) and (b) show the elevated 3D points projected on the 2D SDO/AIA intensity  maps through the LOS technique. LOS1 from Perspective 1 (panel a) is projected down on  Image 2 (panel b). Crosses denote the projection of the elevated coronal points on 2D images. Cross-calibration has been applied for the SDO/AIA and Hinode/EIS images (panels c and d). The Hinode/EIS intensity and Doppler velocity maps match, as long as they are obtained by the same spectrometer at the same time. The LOS projections of the elevated points on the 2D Doppler velocity maps (panels e and f) provide two LOS velocity values attached to their reference point in 3D space. The majority of the reconstructed points are associated with  upflows (blue) and only two points are associated with downflows (red).
		}
		\label{Fig6}
	\end{figure}
	%%%%%%%%%%%%%%%%%%%%%%%%%%%%%%%%%%%%%%%%%%%%%%%%%%%%%%%%%%%%%%%%
	%

We  developed a sequence of algorithms designed to perform triangulation  of Doppler shift  maps of the solar corona recorded  from different  perspectives, with the ultimate goal of enabling reconstruction of the vector velocity field.

Solar-rotation stereoscopy was used to reconstruct the 3D coordinates of the highly elevated points in the solar corona. The same loops were identified in two 2D EUV images in order to calculate the true loop location in 3D corona coordinates. We  used the coronal loops of NOAA  AR 2678 projected in the plane of the sky observed by SDO/AIA  from two different viewing angles (Figure~\ref{Fig6}a,b). The region was stable in time and did not show any flaring activity as demonstrated by 
\href{https://drive.google.com/file/d/1DtK4syJ1oF88HB9eQJKgzIwVx8lb_XIe/view?usp=sharing}{ ({\color{blue} Animation 1})}. The crosses on the images show the different locations of the elevated coronal points when they are projected through the LOS onto 2D images. The SDO/AIA  images and Hinode/EIS intensity images were co-aligned  as described in Section~\ref{STEP2} (Figure~\ref{Fig6}c,d). As long as the Hinode/EIS intensity and velocity maps are co-aligned, 
the crosses on the intensity maps can be directly translated onto the velocity maps. Thus, the crosses on the two Hinode EIS velocity maps indicate two LOS velocities measured in the considered elevated coronal points (Figure~\ref{Fig6}e,f). 
%This assumes that the measured Doppler shift only includes a contribution of the LOS velocity at the 3D point, not of LOS velocity at other points on the (optically thin) LOS.

%%%%%%%%%%%%%%%%%%%%%%%%%%%%%%%%%%%%%%%%%%%%%%%%%%%%%%%%%%%%%%%%%%%

\begin{figure}[ht!]
\centering
{ \includegraphics[width=1.0\hsize]{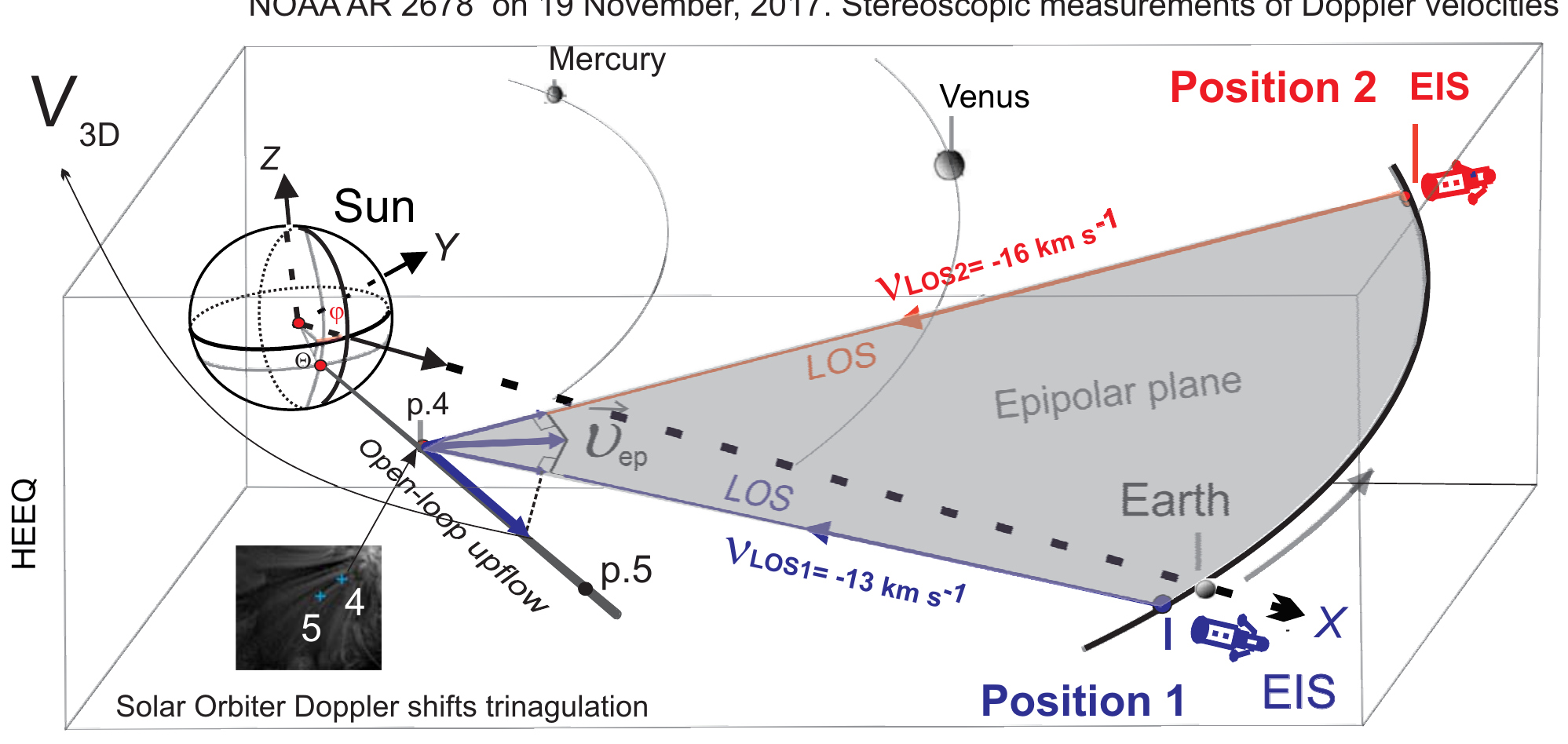}} 
\caption{Stereoscopic observation of NOAA AR 2678 on 19 November 2017, shown in the 3D HEEQ coordinate system. The X axis is directed towards the Earth  and the Z axis is aligned with the solar rotation axis passing through the solar north. The two  observing satellites are separated from each other by 37$\deg$. The blue shifts at point 4 are measured along their respective LOSs and are represented by the vector magnitudes $v_{LOS1}$ and $v_{LOS2}$. The velocity  vector $\vec{v_{ep}}$,  measured in the epipolar  plane, is calculated from $v_{LOS1}$ and $v_{LOS2}$ following the geometry depicted in the figure. In this case, the true velocity vector of point 4 is directed along the coronal loop, showing the plasma flow  confined by the magnetic field, and it is calculated  directly  from the deprojection  of $\vec{V_{3D}}$  on the coronal loop. If the plasma flow is perpendicular  to the epipolar plane, its velocity  cannot be measured, i.e., $\vec{V_{3D}}$ cannot be derived when it makes an angle of 90$\deg$ degrees with  the epipolar plane.}
\label{Fig7}  
\end{figure}
%%%%%%%%%%%%%%%%%%%%%%%%%%%%%%%%%%%%%%%%%%%%%%%%%%%%%%%%%%%%%%%%

\subsection{Velocity vectors for NOAA AR  2678 coronal loops}

In  this  section, we  describe how to  obtain a velocity vector at a point situated high above the photospheric level, using  two LOS Doppler measurements obtained by spectrometers distributed in the heliosphere. Previously, we restored  the true 3D coordinates of points in the high corona and the coordinates of their  projections onto each of the 2D Doppler maps. 

\subsubsection{Deprojection rules of two LOS velocity components in the epipolar plane}

\begin{figure}[!hbt]
			\centerline{
			\includegraphics[width=1.0 \linewidth,clip=]{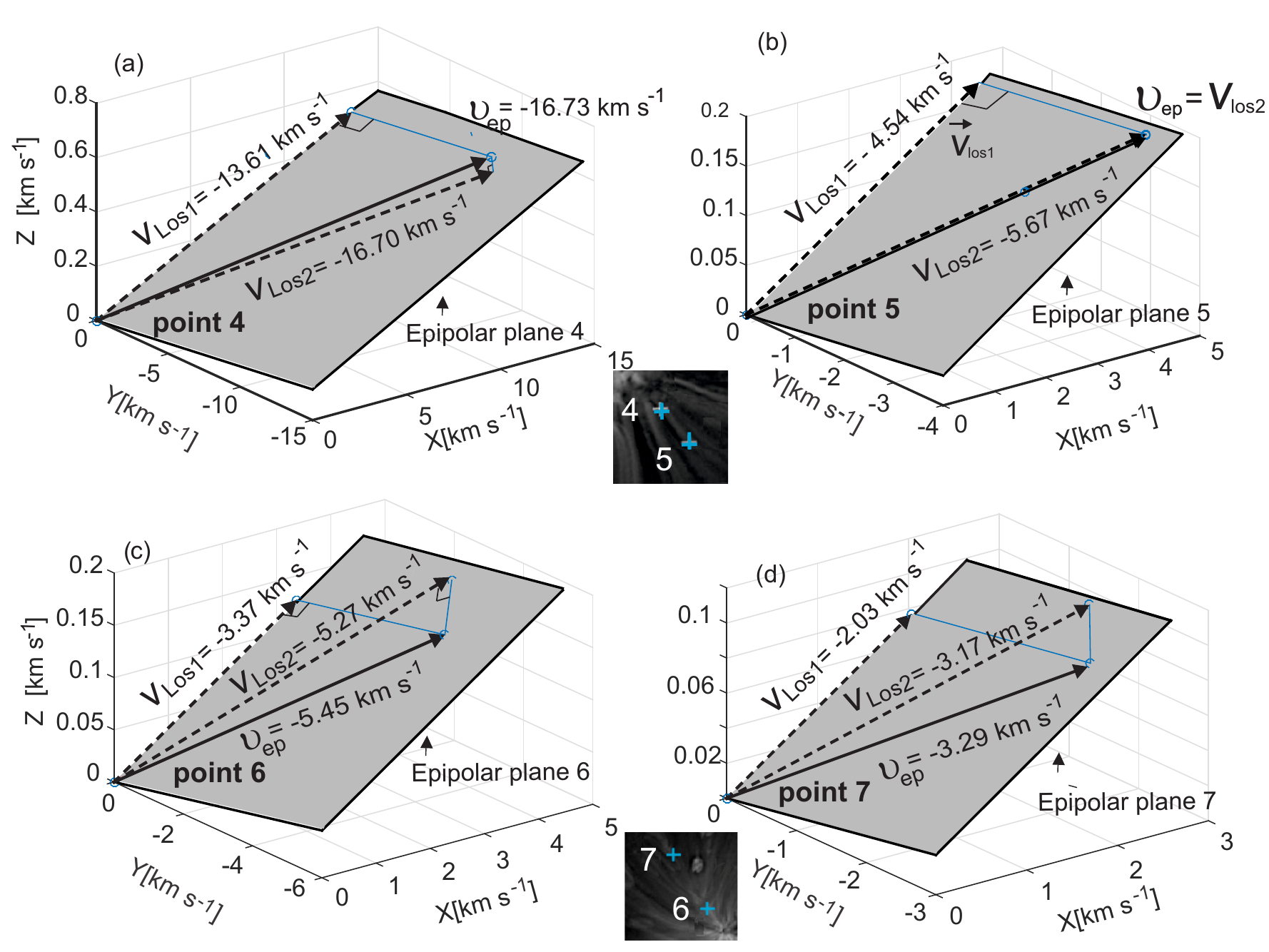}
		}	
		\caption{
		Each plot presents the two LOS velocity components  $\vec{v_{LOS1}}$ and $\vec{v_{LOS2}}$ (dashed arrows), measured by each spectrometer at points 4, 5, 6, and 7 in the corona and the reconstructed $\vec{v_{ep}}$ (solid arrows) in the epipolar plane (gray). The epipolar plane is defined by the locations of the two spacecraft and the point in the solar corona to be triangulated. $\vec{v_{ep}}$  represents the projection  of $\vec{V_{3D}}$ on the epipolar plane. 
		When projected on LOS1 and LOS2, $\vec{v_{ep}}$ represents the velocities $\vec{v_{LOS1}}$ and $\vec{v_{LOS2}}$ measured by each spectrometer.
  }
		\label{Fig8}
	\end{figure}

Observing two  LOS Doppler velocities at any identifiable point  in the corona enables us, first of all, to find the velocity vector in the epipolar plane formed in the 3D heliosphere by the position of the two satellites and the observation point. The gray plane in Figure~\ref{Fig7} shows the epipolar plane defined by the observed point 4  on the open coronal loop and the location of the two satellites. The two Doppler blue-shifted velocities measured at point 4  along the LOSs situated between the observed point and the observing satellites represent the 1D  vectors $\vec v_{LOS1}$ and $\vec v_{LOS2}$. Figure~\ref{Fig8} shows two  LOS Doppler velocities measured at points 4, 5, 6 and 7. Here, $\vec{v_{LOS1}}$  and $\vec{v_{LOS2}}$  are different velocities measured along the LOSs. The velocity vector $\vec{v_{ep}}$  is the projection of the 3D velocity in the epipolar plane, and can be reconstructed by calculating the intersection point of the two perpendiculars deprojected from $\vec{v_{LOS1}}$ and $\vec{v_{LOS2}}$  in the epipolar plane. The velocity vector $\vec{v_{ep}}$,  defined in the epipolar plane, is an especially useful estimate when we cannot identify any structure to which this point belongs. In this case, $\vec{v_{ep}}$  can be an indicator of the plasma flow behaviour.

	%%%%%%%%%%%%%%%%%%%%%%%%%%%%%%%%%%%%%%%%%%%%%%%%%%%%%%%%%%%%%%%%%%%
	%%%%%%%%%%%%%%%%%%%%%%%%%%%%%%%%%%%%%%%%%%%%%%%%%%%%%%%%%%%%%%%%
	
%%%%%%%%%%%%%%%%%%%%%%%%%%%%%%%%%%%%%%%%%%%%%%%%%%%%%%%%%%%

\subsubsection{Plasma outflows along open loops}
\begin{figure*}[ht!]
			\centerline{
			\includegraphics[width=0.9\linewidth,clip=]{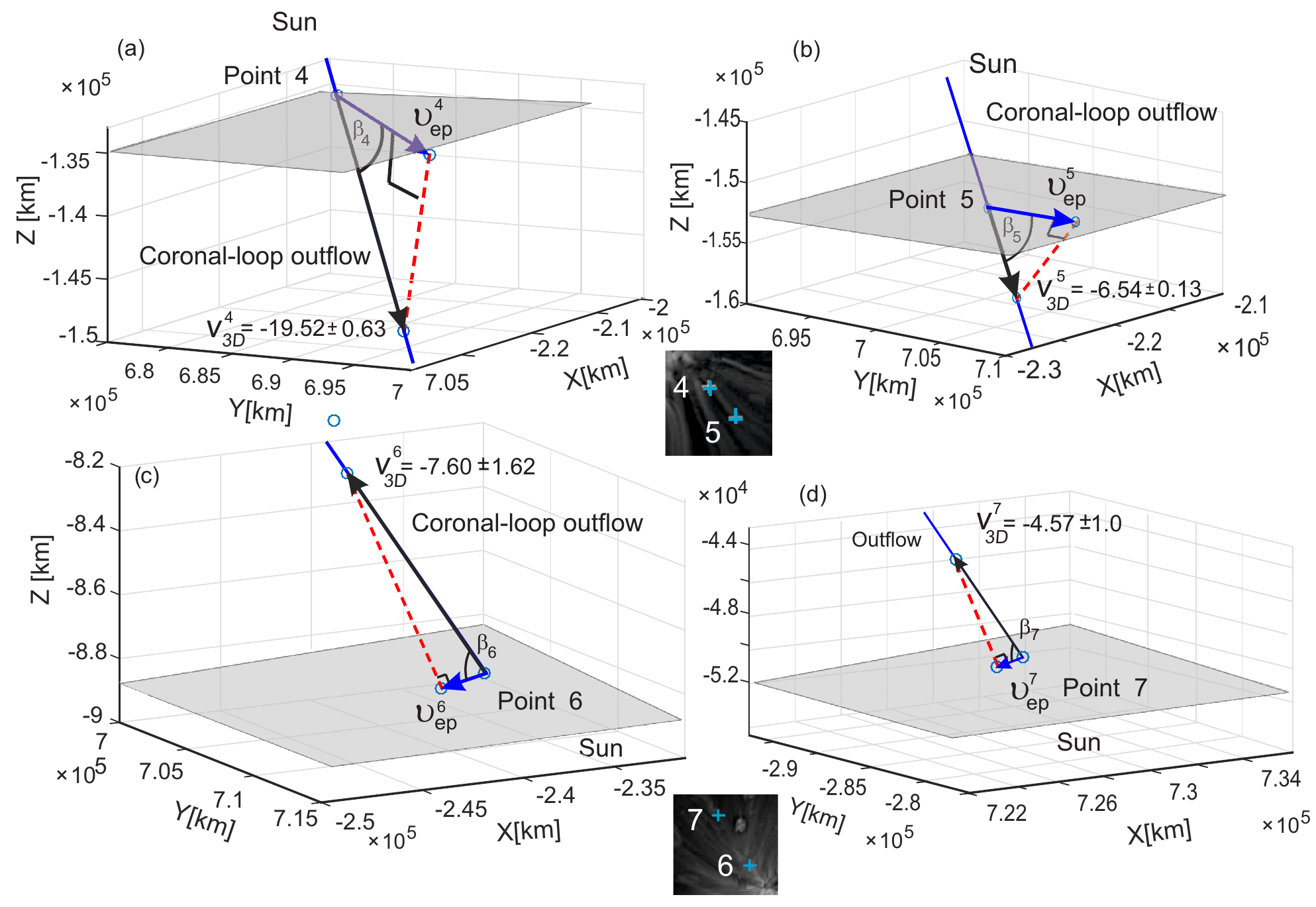}
		}	
		\caption{
		Open loops traced  in  3D solar coronal coordinates at points 4--5 (panels a and b)  and 6--7 (panels c and d).  Gray  planes represent epipolar  planes crossing the loops at  the points  under observation. The epipolar planes are formed in the heliosphere by the two spacecraft locations and each of the points 4, 5, 6, and 7; $\vec{v_{ep}}$ is reconstructed in the plane as shown in Figure \ref{Fig8}.  The $\vec{V_{3D}}$ velocity vectors are reconstructed using the measured angle $\beta$ and $\vec{v_{ep}}$  according to Equation~\ref{coscos}. The magnitudes  of the vectors are shown in Table~\ref{v3dopen}. 
	 }
		\label{Fig9}
	\end{figure*}

In this subsection,  we demonstrate the technique of velocity vector reconstruction in the open coronal loops using the pre-calculated geometry in 3D space. If the pair of points (4, 5)  or (6, 7)  belongs to the open coronal loop approximated as a straight line between two points, we can restore the velocity vector magnitude of the plasma flow by the direct deprojection of  $\vec{v_{ep}}$  on the loop by using the formula:

\begin{equation}
\vec{V_{3D}} = \frac { \vec{v_{ep}} }  { \cos(\beta) } ,
\label{coscos}
\end{equation}
where $\beta$ is the angle between the $\vec{v_{ep}}$ velocity and the coronal loop ( Figure~\ref{Fig9}).  If $\beta=90\deg$ , the deprojection cannot be made. Correspondingly, small $\beta$ angles lead to higher accuracy of the $\vec{V_ {3D}}$ deprojection.

		%%%%%%%%%%%%%%%%%%%%%%%%%%%%%%%%%%%%%%%%%%%%%%%%%%%%%%%%%%%%%%%%%%%
	%%%%%
	
%%%%%%%%%%%%%%%% TABLE 1 
	\begin{table*}[!hbt]
	\centering
		\caption{
		Numerical  values of the fully reconstructed  velocity  vectors $\vec{ V_{3D} }$, which characterise  outflows along open loops shown by black arrows in Figure 9. $\beta$ is the angle between the velocity vector  $\vec{v_{ep}}$ measured in the epipolar plane and the loop. $\vec{V_{3D}}$  can be reconstructed by deprojecting $\vec{v_{ep}}$ onto the coronal loop. The calculations were carried out for each of the points 4, 5,  6, and 7, which lie on open loops. The minus sign is implied to indicate outflows.
}
%\resizebox{\columnwidth}{!}{%
		\begin{tabular} {|c c c c c c |}
	      \hline	
	      	    & $v_{LOS1}$  & $v_{LOS2}$ &  $v_{ep}$ &$\beta$  & $V_{3D}$   \\
	      	   
	      	   & km s$^{-1}$ & km s$^{-1}$ & km s$^{-1}$ & degrees & km s$^{-1}$   \\
	      	  	      \hline
     {\bf Open loop}  &  &  &  &  &    \\
      Point   4 &  -13.61 km s$^{-1}$  & -16.71 km s$^{-1}$ & -16.73 km s$^{-1}$  &28.86$\deg$ & -19.11 km s$^{-1}$   \\

         Point   5 &  -4.54 km s$^{-1}$  & -5.67 km s$^{-1}$  & -5.67  km s$^{-1}$ & 28.54$\deg$ & -6.45 km s$^{-1}$     \\

    {\bf Open loop}  &  &  &  &  &    \\
    Point 6    &  -3.37  km s$^{-1}$  & -5.27  km s$^{-1}$ & -5.45  km s$^{-1}$ & 36.29$\deg$  &   -6.76 km s$^{-1}$   \\

       Point 7     &  -2.03 km s$^{-1}$  & -3.17 km s$^{-1}$  & -3.29 km s$^{-1}$  & 36.32$\deg$ & -4.07 km s$^{-1}$  \\
        
            \hline
          \end{tabular}
          \label{v3dopen}
   %       }
		\end{table*}
		%%%%%%%%%%%%%%%%% TABLE 1 <

We  reconstructed the velocity  vector $\vec{V_{3D}}$  for the points in the open-loop outflows using  deprojections:  $\vec{v_{ep}} \rightarrow \vec{V_{3D}}$. Figure~\ref{Fig9} shows the locations of the coronal loop in 3D corona coordinates and the epipolar planes crossing loops at the points under observation, where two LOS Doppler
velocities are measured and $\vec{v_{ep}}$ is  obtained (blue arrows in the planes). After the measurement of the $\beta$ angle  between $\vec{v_{ep}}$ (blue arrows) and the loop in 3D space, we obtain $\vec{V_{3D}}$ (black arrows along the loops) using Equation~\ref{coscos}. In all cases, the resulting vector velocities $\vec{V_{3D}}$ are directed outwards. The numerical values of the velocity vectors are given in Table~\ref{v3dopen}. The small variation in the angle $\beta$ between the LOS Doppler velocities (same spacecraft position) and the open loop  when measured from points on the same loop reflects the location change of the observation point.

\subsubsection{Plasma outflows in closed loops}

\begin{figure}[ht!]
			\centerline{
			\includegraphics[width=1.0 \linewidth,clip=]{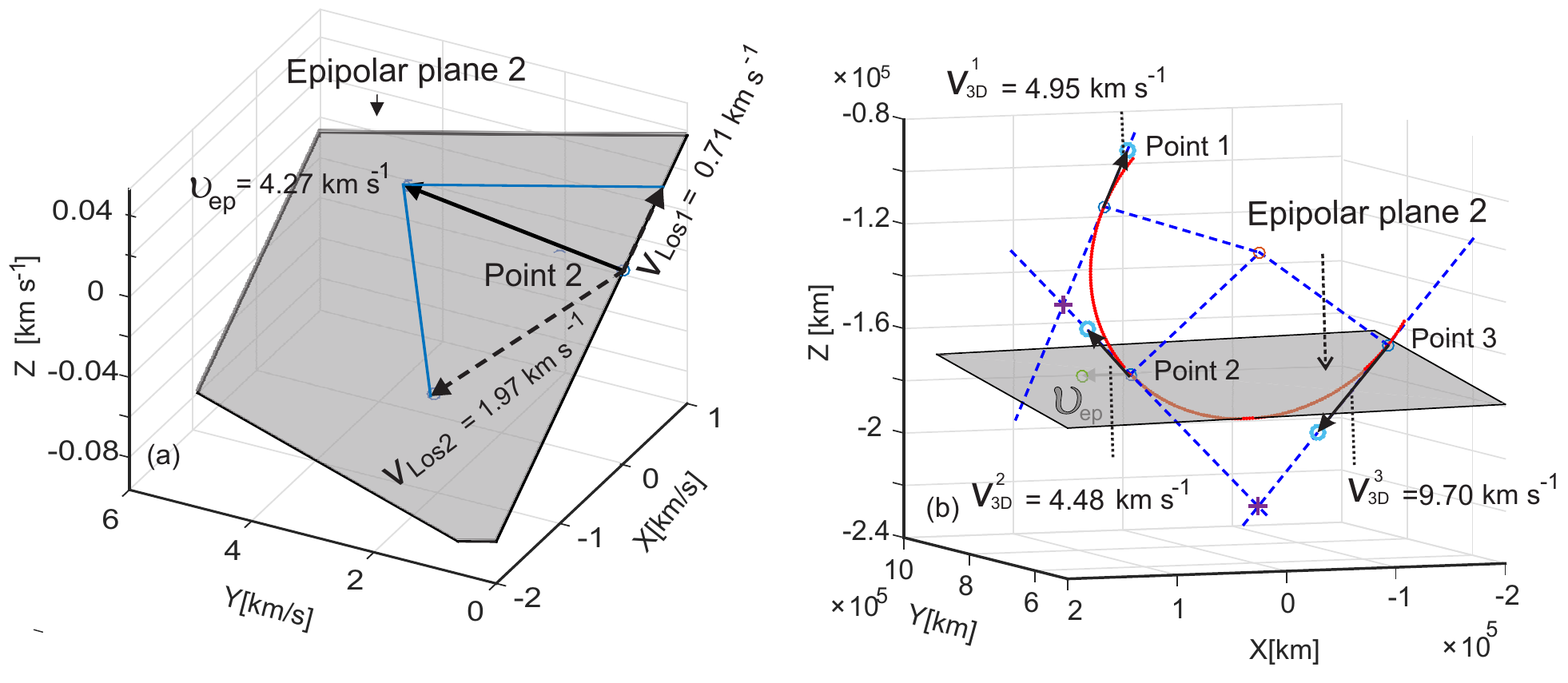}
		}	
		\caption{ 
Plasma flow velocity  vectors in the 3D closed coronal loop under investigation. The location of the loop points in the high corona and the  3D geometry of the loop were determined using the LOS projection technique. (a) shows the velocity vector $\vec{v_ {ep}}$, deprojected from the two Doppler  LOS velocities measured by the two spectrometers at point 2.  The epipolar plane (gray)  is  the plane containing  the two  satellites and the observation  point. The vectors $\vec{V_{3D}}$ were constructed by the deprojection  of $\vec{v_ {ep}}$ on the tangent to  the loop direction  (panel b). 
%The one gray  arrow  at  point 2 is $\vec{v_ {ep}}$, reconstructed  by the deprojection shown in the left panel 
The absolute values of the vectors are shown in Table~\ref{v3dclosed}.
}
		\label{Fig10}
	\end{figure}

In this subsection, we demonstrate a technique for determining the velocity vector for flows following curved paths traceable in the solar corona. When measuring the LOS Doppler velocities with two spectrometers at points belonging to a closed, semicircular or simply curved,  coronal structure, $\vec{V_{3D}}$   can be determined again using Equation~\ref{coscos}. However, in this case, the deprojection should be performed on the tangent  to the coronal-loop direction, as long as the velocity vector is oriented in  this  direction. Figure~\ref{Fig10} (left  panel)  shows the deprojected velocity 
$\vec{v_ {ep}}$ from the two measured LOS Doppler velocities $\vec{v_{LOS1}}$, $\vec{v_{LOS2}}$. Figure~\ref{Fig10} (right panel) shows $\vec{V_ {3D}}$  at points 1, 2, and  3.  Detailed information on the LOS Doppler velocity magnitudes and their angles with the tangents are listed in Table~\ref{v3dclosed}.

%
	%%%%%%%%%%%%%%%%%%%%%%%%%%%%%%%%%%%%%%%%%%%%%%%%%%%%%%%%%%%%%%%%%%%
	%%%%%%%%%%%%%%%%%%%%%%%%%%%%%%%%%%%%%%%%%%%%%%%%%%%%%%%%%%%%%%%%
%

%%%%%%%%%%%%%%%% TABLE 1 
	\begin{table*}[!hbt]
	\centering
		\caption{
		Velocity  vectors characterising  the flows in  the investigated closed loops observed in NOAA AR  2678 on November 2017. $\vec{V_{3D}}$,  shown in Figure~\ref{Fig10}, is reconstructed at points  1,  2, and  3 on the loop.  The velocity  vectors $V_{3D}$  are found by deprojection  from  $\vec{v_{ep}}$ on the tangent to the coronal loop directions. The minus sign is implied to indicate outflows.}
		
%	\resizebox{\columnwidth}{!}{%
	\begin{tabular} {|c c c c c c |}
 %{| p{1.3cm} p{2.3cm} p{2.3cm} p{2.3cm} p{1.0cm} p{0.7cm}|}
	     	      \hline	
	      	    & $v_{LOS1}$  & $v_{LOS2}$ &  $v_{ep}$ & $\beta$  & $V_{3D}$      \\
	      	   
	      	  & km s$^{-1}$ & km s$^{-1}$ & km s$^{-1}$ & degrees & km s$^{-1}$  \\
	      	  	      \hline
 {\bf Closed loop}  &  &  &  &  &    \\	      	  	      
   Point 1    &  2.38 km s$^{-1}$  & -0.68 km s$^{-1}$  & 4.91 km s$^{-1}$  & 7.51$\deg$  &   4.95 km s$^{-1}$ \\

  Point 2       &  0.71 km s$^{-1}$  & -1.97 km s$^{-1}$  & 4.27 km s$^{-1}$  &  17.98$\deg$ & 4.48 km s$^{-1}$   \\

    Point 3   &  -7.35 km s$^{-1}$  & -9.4 km s$^{-1}$  & -9.41 km s$^{-1}$  & 14.11$\deg$ &  -9.70 km s$^{-1}$   \\

       \hline
          \end{tabular}
          \label{v3dclosed}
 %         }
		\end{table*}
		%%%%%%%%%%%%%%%%% TABLE 1 <

Thus, the velocities deprojected from two LOS Doppler velocities measured at points of a closed coronal loop provide true intrinsic information of the plasma flow behaviour in the loop. In this case, the deprojection is not carried out on the structure, as described in the previous subsection, but on the direction of the tangent to the loop reconstructed  at the observed  point.  This technique can be applied to any curved structure in the solar atmosphere, provided that  we can define its  3D geometry using, for instance, epipolar geometry stereoscopy or magnetic modelling of the loops.

\subsubsection{Comparison of deduced inclinations and linear force-free field extrapolation of NOAA AR 2678}
\label{B_model}

Magnetic field modelling has been used extensively  to interpret the morphology of coronal magnetic loops \citep{Chifu2015}. An example by \citet{harra_2008} illustrates how using the angles from magnetic field modelling enables a more accurate determination of the upflowing plasma at the edge of an active region. 

We compute the coronal magnetic field topology of NOAA AR 2678 during its disc transit on 11 November 2017 at 08:28 UT and 22 November 2017 at 03:44 UT. The LOS magnetic field is extrapolated to the corona using a linear force-free field (LFFF) configuration where $\vec{J} \times \vec{B} =0 $ and $\nabla \times \vec{B} = \alpha_{\rm M} \vec{B}$, with $\alpha_{\rm M}$ constant \citep{Mandrini1996,Demoulin1997}.

%%%%%%%%%%%%%%%%%%%%%%%%%%%%%%%%%%%%%%%%%%%%%%%%%%%%%%%%%%%%%%%%%%%%%%
	\begin{figure}[!hbt]
				\centerline{
			\includegraphics[width=1.0 \linewidth,clip=]{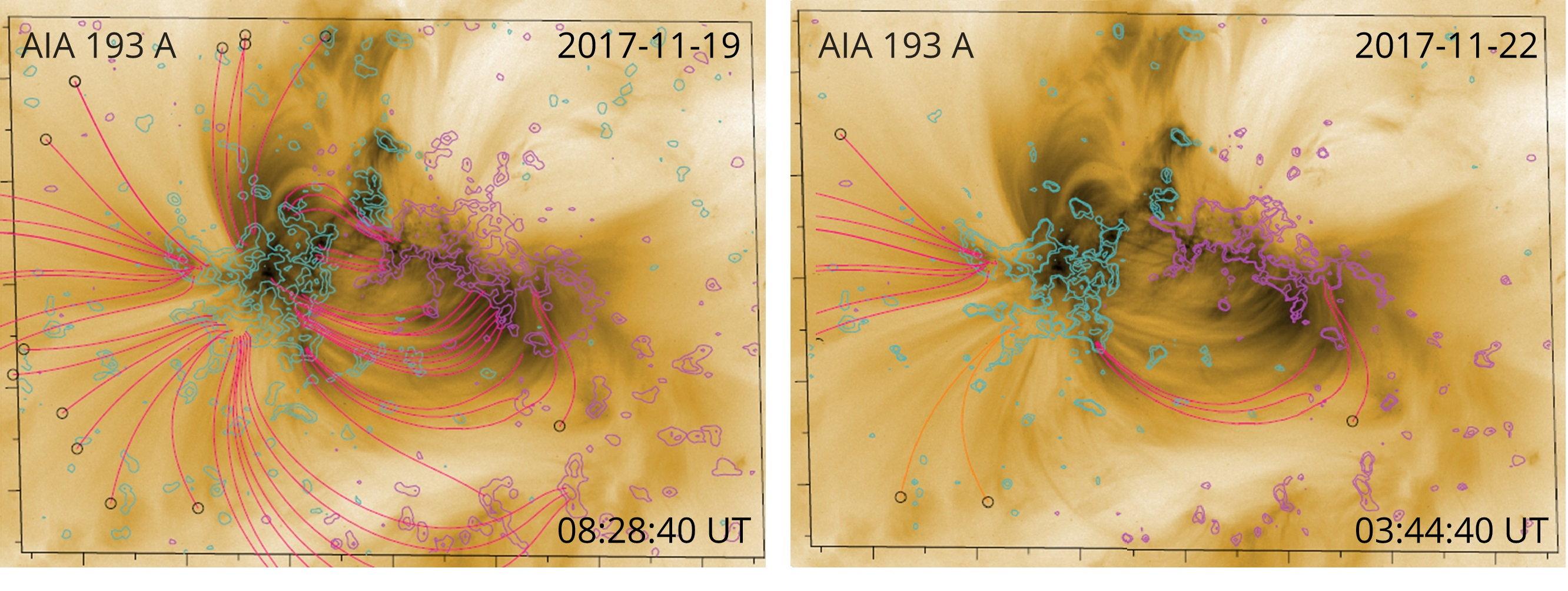}
		}	
		\caption{LFFF extrapolation for NOAA AR  2678 overlaid on an SDO/AIA intensity map. The left panel shows the LFFF model results at 08:28 UT  on 19 November 2017. The right panel shows the LFFF model extrapolation at 03:44 UT on 22 November 2017.}
		\label{Fig11}
	\end{figure}
	%%%%%%%%%%%%%%%%%%%%%%%%%%%%%%%%%%%%%%%%%%%%%%%%%%%%%%%%%%%%%%%%%%%%%%
	
%	%%%%%%%%%%%%%%%%%%%%%%%%%%%%%%%%%%%%%%%%%%%%%%%%%%%%%%%%%%%%%%%%%%%
%	\begin{figure}[!hbt]
% 			\centerline{
%			\includegraphics[width=0.9 \linewidth,clip=]{Figures/Fig11.pdf}
%		}	
%		\caption{ 
%		LFFF extrapolation for NOAA AR  2678 overlaid on an SDO/AIA intensity map. The left panel shows the LFFF model results at 08:28 UT  on 19 November 2017. The right panel shows the LFFF model extrapolation at 03:44 UT on 22 November 2017.}
%		\label{Fig11}
%	\end{figure}
	%%%%%%%%%%%%%%%%%%%%%%%%%%%%%%%%%%%%%%%%%%%%%%%%%%%%%%%%%%%%%%%%

Figure~\ref{Fig11} (left panel), shows the LFFF model results for 19 November 2017. Similar results are obtained on 22 November 2017 (right panel). Both panels show the modelled global coronal structure for which we have used the closest in time magnetograms to each EIS image as boundary condition.  The panels are constructed superposing field lines computed using different values of $\alpha_{\rm M}$, which have been selected to better match the shape of the observed loops in AIA 193~\AA. To do this comparison, the model is first transformed from the local frame to the observed frame as discussed in \citet{Mandrini2015} \citep[see the transformation equations in the Appendix of][]{Demoulin1997}. This allows a direct comparison of our computed coronal field configuration to AIA EUV lopps. Furthermore, in order to determine the best matching $\alpha_{\rm M}$ values we have followed the procedure discussed by \citet{Green2002}. 

\begin{figure}[!hbt]
			\centerline{
			\includegraphics[width=0.7 \linewidth,clip=]{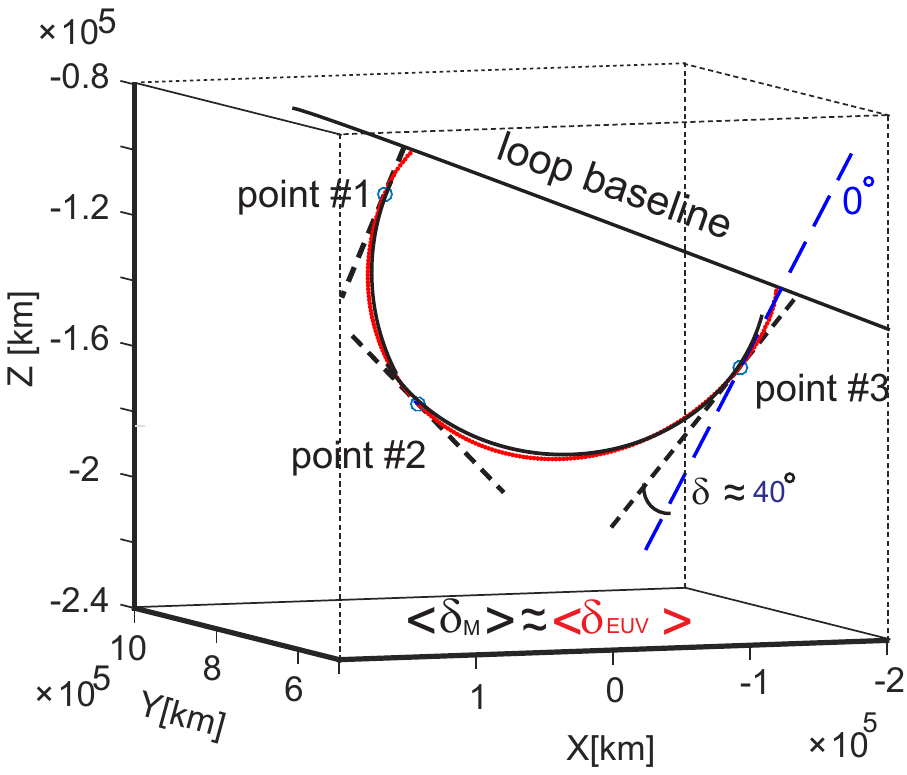}
		}	
		\caption{ 
		Closed coronal loop of NOAA AR  2678 traced  in 3D corona coordinates by stereoscopic (red) and LFFF (black) methods.  We computed here the inclination $\delta_{EUV}$ between the vertical  to the solar surface (blue) and the tangential line at the studied points for SDO/AIA loops using stereoscopic methods and we found a close similarity with  the $\delta_M$  computed using the LFFF extrapolations.
		}
		\label{Fig12}
	\end{figure}

Figure~\ref{Fig12} shows the definition of the loop inclination angles $\delta_{EUV}$ computed for EUV SDO/AIA loops using triangulation methods, which  are compared with those $\delta_M$  derived from our LFFF extrapolation. The inclination angles $\delta_M$ derived from the model are computed as discussed in \citet{Demoulin2013}; they are the angles between the line tangential to the loop at the coronal points studied here and the vertical to the solar surface as shown on the Figure~\ref{Fig12}.

	%%%%%%%%%%%%%%%%%%%%%%%%%%%%%%%%%%%%%%%%%%%%%%%%%%%%%%%%%%%%%%%%%%%
	
	%%%%%%%%%%%%%%%%%%%%%%%%%%%%%%%%%%%%%%%%%%%%%%%%%%%%%%%%%%%%%%%%
	
Both LFFF  and stereoscopic methods obtained rather close ranges of inclination angles to the vertical:  $[20\deg, 30\deg]$ at point 1 and $[35\deg, 45\deg]$  at point 3  on the closed loop. 
For open loops, the inclination $\delta_{EUV}$ and $\delta_M$ to the vertical at point 6 and point 7 (where the field lines are quite simple) are in the same range $[10\deg, 30\deg]$.  At point 4 and point 5 the inclination angles estimated by both methods are in the range between $[15\deg, 25\deg]$. Since the projection of an observed coronal loop has a certain thickness, a set of computed infinitely thin field lines can match the visible shape of a loop; this is the origin of the range of values for the inclination $\delta_M$ coming from the magnetic field model.

In general, we do not observe discrepancies between the field lines computed from the LFFF model and stereoscopic reconstruction, although photospheric footpoints are difficult  to identify in EUV images and from the magnetic model we trace the full line starting at the photospheric level. The latter demonstrates the future  opportunity of pairing stereoscopic loop reconstruction using Solar Orbiter data with magnetic field  modelling, which is expected to  significantly improve  the accuracy of stereoscopic vector velocity measurements   (see, {\it e.g.}, \citealt{Aschwanden2015}).

\section{3D Spectroscopy by two distributed  imaging spectrometers}
\label{STEP5}

We developed an approach using a pair of space-born spectrometers that provide stereoscopic views of Doppler shift maps in the solar corona. Measurement of the velocity vector in the corona can be achieved with one extraterrestrial remote spectrometer (SO/SPICE) and with the orbital spectrometer Hinode/EIS or IRIS. Dual spectrometers allow measurement of the 2D velocity vector in any identifiable coronal point, and eventually reconstruction of 3D velocity vectors, using the morphology of the coronal structures. In this section we describe how we derive  the optimal spacecraft configurations for 3D spectroscopy with SO. The spectral and spatial resolution of spectrometers influences directly the velocity vector measurement  accuracy.
 
\subsection{SPICE, Hinode/EIS  and IRIS spectrometers}

High-resolution spectroscopy of optically  thin plasma allows measurements of plasma parameters such as temperatures, densities, chemical abundances, and Doppler and non-thermal motions \citep{Phillips2012,DelZanna2018}. 

In the science phase of the Solar Orbiter mission, there will be opportunities to carry out 3D spectroscopy in the chromosphere, transition region, and corona for the first time by combining data from Hinode/EIS \citep{EIS}, IRIS \citep{IRIS} and SO/SPICE \citep{AndersonSPICE2019}. Hinode/EIS wavelength range is 1170-210~\AA\ and 250-290~\AA, and its angular resolution is 2 arcsec. There are four slit and slot positions: 1 arcsec slit,  2 arcsec slit,  40 arcsec slot, and 266 arcsec slot. The temporal resolution is a few seconds in dynamic events, 10 s in active regions and around 1 minute in coronal holes. The maximum FOV is 360 arcsec $\times$ 512 arcsec. 

The SO/SPICE instrument is also a high-resolution imaging spectrometer operating in EUV/UV wavelengths. Its design is focused on studies that combine remote sensing and in situ instruments on board Solar Orbiter. The wavelength range is 704--790~\AA\ and 973--1049~\AA\ and the accuracy of the line shifts is ~5 km/s. Table~\ref{table:eis} shows the different emission lines used by SPICE, their wavelengths, temperature of formation and
their closest equivalents for Hinode/EIS. In most cases, the emission lines for the two instruments do not represent the same ion, but the formation temperature is similar. Some caution must be used for lines that are optically thick, such as He {\sc ii}, when comparing them.

IRIS is an imaging spectrometer with  a slit-jaw imaging system that probes the chromosphere. It explores the solar chromospheric dynamics to determine how the energy flows through the chromosphere and the transition region. The spacecraft has a polar sun-synchronous orbit.  IRIS has a spatial resolution of 0.33–0.4 arcsec, a temporal resolution of 2 s, and a velocity resolution of 1 km/s over an FOV of up to 175 arcsec $\times$ 175 arcsec \citep{bart,Bart2014}. 
 
\begin{table}[!hbt]
\caption{SPICE emission lines, wavelength, temperature  of formation (T),  and the closest equivalents in Hinode/EIS.}
\resizebox{\columnwidth}{!}{%
\begin{tabular}{|ccc|ccc|}
\hline
SPICE  & Wavelength (\AA)  & $\log T$ & EIS  & Wavelength (\AA) & $\log T$ \\
\hline
C~{\sc iii} & 977.03 & 4.5 & He~{\sc ii} & 256.7 & 4.7 \\

O~{\sc v} & 760.43 & 5.4 & O~{\sc v} & 192.9 & 5.4 \\

Ne~{\sc viii} & 770.92 & 5.8 & Mg~{\sc vii} & 278.39, 280.75 & 5.8 \\
  	& 	& 	& Si~{\sc vii} & 275.35 & 5.8 \\
	&	&	& Fe~{\sc viii} &185.21, 194.66 & 5.8 \\

Mg~{\sc ix} & 706.02 & 6.0 & Fe~{\sc x} & 184.5, 257.2 & 6.0 \\
Fe~{\sc x} & 1028.04 & 6.0 & & & \\

Mg~{\sc xi} & 997.44 & 6.2 &  Fe~{\sc xiii} & 196.54, 202.04, 203.83 & 6.2 \\

Si~{\sc xii} & 520.67 & 6.3 & Fe~{\sc xiv} & 274.20, 264.79 & 6.2 \\
\hline
\multicolumn{6}{|c|}{Flare lines} \\
\hline

Fe~{\sc xviii} & 974.84 & 6.9 & Ca~{\sc xvii} & 192.82 & 6.7 \\
Fe~{\sc xx} & 721.55 & 7.0 & Fe~{\sc xxiii} & 263.76 & 7.2 \\
	& 	& 	& Fe~{\sc xxiv} & 255.11 & 7.2 \\
\hline	
\end{tabular}
\label{table:eis}
}
\end{table}

IRIS and Hinode/EIS frequently carry out simultaneous observations, pointing at the same target. This mode of collaboration will be of great interest alongside the different FOVs of Solar Orbiter. The IRIS wavelength range covers the chromosphere, and the Hinode/EIS wavelength range covers the corona. The SPICE wavelength range provides measurements of the chromosphere, transition region, and corona. Combining the three spectrometers will provide the first 3D solar spectroscopy. Table~\ref{table:iris} shows a comparison of the SPICE and IRIS wavelengths.  

\begin{table}[!hbt]
\caption{SPICE emission lines, wavelength, temperature  of formation, and the closest equivalents in IRIS.}
\resizebox{\columnwidth}{!}{%
\begin{tabular}{|ccc|ccc|}
\hline
SPICE  & Wavelength  (\AA)  & $\log T$ & IRIS  & Wavelength (\AA) & $\log T$ \\
\hline
H~{\sc i} & 1025.72 & 4.0 & Mg~{\sc ii} & 803, 2796 & 3-7--3.9 \\

C~{\sc ii} & 1036.34 & 4.3 & C~{\sc ii} & 1334, 1335 & 4.3 \\

O~{\sc iv} & 787.72 & 5.2 & O~{\sc iv} & 1399, 1401 & 5.2 \\

Mg~{\sc xi} & 997.44 & 6.2 & Fe~{\sc xii} & 1349.4 & 6.2 \\

\hline
\multicolumn{6}{|c|}{Flare lines} \\
\hline

Fe~{\sc xx} & 721.55 & 7.0 & Fe~{\sc xxi} & 1354.1 & 7.0 \\

\hline
\end{tabular}
\label{table:iris}
}
\end{table}

\subsection{SPICE observing with  the Extreme Ultraviolet  Imager and SDO/AIA}
\label{subsec:3D_eui_aia}

For 3D geometry reconstruction, we used SDO/AIA  intensity images and their coupling with Doppler velocity maps
to perform operations with vector velocities. In this context, Solar Orbiter Doppler maps can be combined with the intensity
images of the Extreme Ultraviolet  Imager (EUI)  on board the same satellite. In both cases, the match between Doppler maps and intensity images provides all the data necessary to compute velocity vectors.
%for vector operations with the velocities.

\subsubsection{The EUI}
The Extreme Ultraviolet Imager (EUI, \citealt{RochusEUI2019}), on board  Solar Orbiter observes the solar atmosphere from the top of the chromosphere to the low corona using three imaging telescopes: the Full Sun Imager (FSI) and two High Resolution Imagers (\hrieuv~and \hrilya). The FSI provides a 3.8$\deg$ $\times$ 3.8$\deg$  FOV in the 174~\AA\ and 304~\AA\ passbands with a typical temporal cadence of 600 s, and a spatial resolution of 9 arcsec. This telescope provides a context view of the upper solar atmosphere up to $4 R_{\sun}$ at perihelion.

Both \hrieuv~and \hrilya~telescopes  provide a 17 arcmin $\times$ 17 arcmin FOV  with  a plate-scale of 0.5 arcsec and unprecedented  imaging cadence of 1 s. The passband of \hrieuv~(174~\AA) is comparable to that of FSI. The second high-resolution telescope, \hrilya, observes the Sun in the Lyman $\alpha$ line (Ly$\alpha$\, 1216~\AA). The HRI instruments will study the small-scale structures and highly dynamic events in the upper solar atmosphere. Ly$\alpha$ is the most intense emission line in the solar spectrum and affects planetary atmospheres, so it is of great interest; however, so far observations in Ly$\alpha$ are rare and were carried out mainly aboard sounding rockets  \citep{Korendyke2001,Vourlidas2010,Vourlidas2016, Chua2013,Chintzoglou2018} and the Transition Region and Coronal Explorer space observatory \citep{Golub1999}. Table 6 shows the closest corresponding SPICE emission line to the EUI passband. Again, caution must be used for the temperature of formation with Ly$\alpha$, as it is a complex, optically thick spectral line.

\begin{table}[!hbt]
\caption{SPICE  emission lines, wavelength, and temperature of formation,  and the closest equivalent filter  in EUI.}
\resizebox{\columnwidth}{!}{%
\begin{tabular}{|ccc|ccc|}
\hline
SPICE  & Wavelength (\AA)  & $\log T$ & EUI  & Wavelength (\AA) & $\log T$ \\
\hline

C~{\sc iii} & 977.03 & 4.5 & Ly$\alpha$ & 1216 & 4.7 \\

Mg~{\sc viii} & 772.31 & 5.9 & Fe~{\sc ix} & 174 & 5.9 \\

Ne~{\sc viii} & 772.31 & 5.9 & Fe~{\sc ix} & 174 & 5.9 \\
\hline
\end{tabular}
\label{table:eui}
}
\end{table}

\subsection{3D-point triangulation uncertainties}
\label{geom_err}

\subsubsection{Spacecraft separation angle and spatial resolution}
An issue of great importance when planning observations with two spacecraft is understanding the optimum separation between them. In this section, we discuss how the accuracy of the measurement  varies with spacecraft separation.%\cmc{Check this there are two definitions for $ds$} 
\citet{Inhester2006} calculated that the errors in 3D coronal point triangulation  depend on both the spatial resolution $ds$ of an image and on the spacecraft separation angle $\gamma$:
\begin{equation}
\varepsilon_{\Delta} = \frac{ds}{\sin(\gamma/2)},
\label{Err_tri}
\end{equation}
$ds$ is the pointing error across the epipolar line, which is linearly dependent on the spatial resolution of the instrument and can have values from 1 to 10 pixels, depending on with what accuracy  one can identify the same feature in two images. When the spatial resolution is fixed and $\gamma = 0$, the errors are very large. For large $\gamma$ the errors are small, but additional errors are introduced due  to complications in identifying the same feature in both images. When $\gamma = 180\deg$, the errors are very large for off-limb features because  the LOSs of the two  spacecraft are parallel. The low optical thickness of the solar corona enables the observation of the same off-limb feature when the separation is near 180$\deg$, but makes it impossible  to observe  the same on-disk feature \citep{Aschwanden2012b,Aschwanden2015}.

	%%%%%%%%%%%%%%%%%%%%%%%%%%%%%%%%%%%%%%%%%%%%%%%%%%%%%%%%%%%%%%%%%%%
	\begin{figure}[!hbt]
			\centerline{
			\includegraphics[width=0.8 \linewidth,clip=]{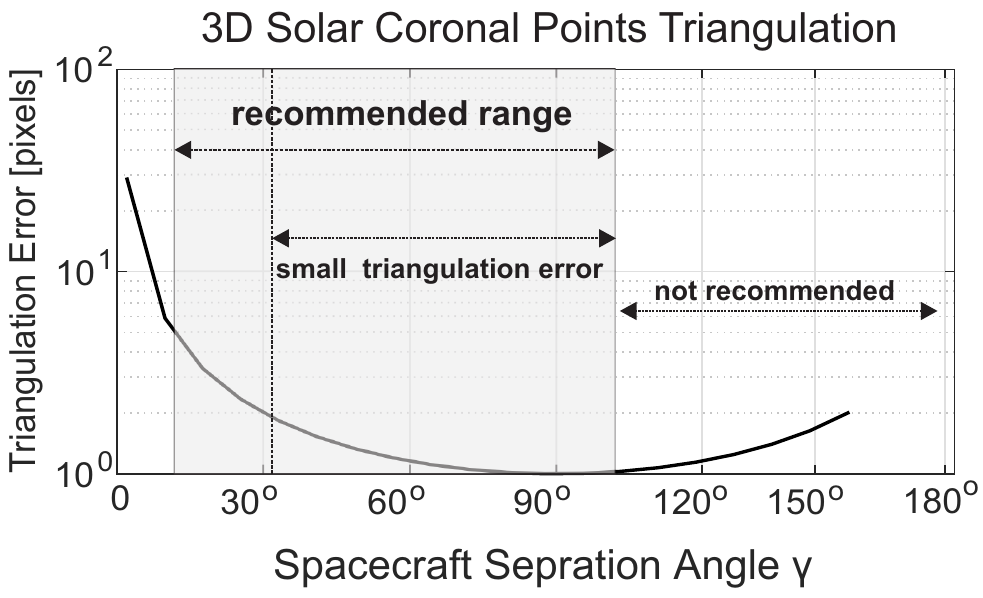}
		}	
		\caption{
		Geometric  error  for  triangulating a single-point  object  as a function  of spacecraft separation  angle. The  object  is observed by  two  separated spacecraft,  e.g.,  the Solar Orbiter and one Earth-orbiting spacecraft (SDO,  Hinode,  or  IRIS). The separation angles between the Earth–Sun line and the Solar Orbiter-Sun line varies in the all range of angles during the cruise and nominal phase of the mission.  Angles of $\pm90\deg$  provide the minimal triangulation error of 1.4 pixels for the 3D geometry reconstruction  using Solar Orbiter high resolution imagers \hrieuv~and \hrilya. Separation angles greater than 100$\deg$ between spacecraft can complicate pointing at the same area of the Sun with two spectrometers and is therefore not recommended.. However, the angle interval of $10\deg$ -- $100\deg$  provides a sufficiently small triangulation error and good pointing possibilities for spectrometers.
	}
		\label{Fig13}
	\end{figure}
	%%%%%%%%%%%%%%%%%%%%%%%%%%%%%%%%%%%%%%%%%%%%%%%%%%%%%%%%%%%%%%%%

The geometric error in Equation~\ref{Err_tri} is the error for the triangulation of a single-point object.  For a 1D object such as  a section of a loop, the error is more complex, as it additionally increases with decreasing angle between the loop tangent and the epipolar plane, and theoretically becomes infinite if this angle becomes zero. The intuitive reason for this behaviour is that a smaller angle with the epipolar plane results in a smaller visual angle between the projected tangent and the epipolar lines in both images. Consequently, the point of intersection of the loop with the epipolar lines can be less well determined \citep{Inhester2006}.

\citet{Aschwanden2012b}  showed that stereoscopy of coronal loops is feasible with good accuracy from around $6\deg$  to $127\deg$. This range is also predicted theoretically by taking into account  the triangulation  errors due to finite spatial resolution and uncertainty  in the identification of stereoscopic correspondence in image pairs, which is disturbed by projection effects and foreshortening for viewing angles near the limb. By combining the accuracy of altitude triangulation with the stereoscopic correspondence ambiguity, the authors estimated that a spacecraft separation angle of  $22\deg$ -- $125\deg$  is most favourable for stereoscopy, when using an instrument with a spatial resolution of 2.6 arcsec (pixel size of 1.6 arcsec) such as Extreme Ulltaviolet Imager aboard STEREO.

We used Equation 2 to calculate the triangulation errors for different spacecraft separation angles $\gamma$ between the  Solar Orbiter and the near-Earth spacecraft Hinode or IRIS. Figure~\ref{Fig13} shows the triangulation error as a function of spacecraft separation angle, where $ds$ is 1 pixel. The physical measure  contained in a 1-pixel unit will vary as (100-350)~km$^2$ for HRI and as (900--3000)~km$^2$  for FSI when the Solar Orbiter travels between aphelion and perihelion. The angle interval of $10\deg$ -- $100\deg$  is expected to provide a sufficiently small triangulation error and good pointing possibilities for 3D solar spectroscopy. 

%%%%%%%%%%%%%%%%%%%%%%%%%%%%%%%%%%%%%%%%%%%%%%%%%%%%%%%%%%%%%%%%%%%%
\section{Discussion and conclusions}
\label{sec:Discussion}
%%%%%%%%%%%%%%%%%%%%%%%%%%%%%%%%%%%%%%%%%%%%%%%%%%%%%%%%%%%%%%%%%%%%

We have developed a technique that reconstructs velocity vectors  in coronal loops from the Doppler maps and EUV images obtained  by  two  spacecraft  located  at  significant  (on  a  heliospheric scale) distances from each other. 
The methodology is built upon the dynamic solar rotation EUV spectroscopy and on STEREO triangulation methods for coronal loops observed in optically thin coronal EUV emissions lines \citep{Aschwanden2011b}. Novel algorithms use methods of spatial analytical geometry in order to determine 3D velocities in coronal loops using different LOSs Doppler shifts measurements. The study is inspired by the launch of the Solar Orbiter in February 2020, the first solar mission to travel far beyond the ecliptic plane, which carries the EUV spectrometer. For the first time, stereoscopic pairs of solar Doppler maps will be available by combining data from the distant Solar Orbiter and the near-Earth spectrometers. We, therefore, explore the optimum conditions for future vector velocity observations with Solar Orbiter. The methodology consists of the following steps:

\begin{enumerate}
\item {\bf Determination of 3D  geometry} of structures through  the LOS projection  technique with high resolution broad band EUV imagers aboard Solar Orbiter EUI, Solar Dynamics Observatory and STEREO. 
\item {\bf Cross-calibration  and alignment}  between monochromatic and broad--band EUV images.
\item {\bf Vector velocity reconstruction}
 \begin{enumerate}
\item \underline{Data acquisition}: Stereoscopic pair of Doppler shifts at each coronal point.
\item \underline{Deprojection "1D$\rightarrow$2D"}:
$\vec{v_{LOS1}}$,$\vec{v_{LOS1}} \rightarrow $ $\vec{v_{ep}}$. Two measured LOS velocity components are deprojected into velocity vector $\vec{v_{ep}}$ defined in the plane formed by the two spacecrafts and the coronal point we reconctruct the velocity vector. When orthogonally projected on LOS1 and LOS2, $\vec{v_{ep}}$ represents $\vec{v_{LOS1}}$ and $\vec{v_{LOS2}}$ velocities measured by each spectrometer.   
\item \underline{Deprojection "2D$\rightarrow$3D"}:  $\vec{v_{ep}} \rightarrow \vec{V_{3D}}$. The velocity vector $\vec{V_{3D}}$  is reconstructed  by the orthogonal deprojection of the defined in the observational plane $\vec{v_{ep}}$ onto the direction of plasma flow in the 3D solar corona.
\item \underline{Limits of application}: If the plasma flow in the solar corona is orthogonal to the observational plane  the velocity vector cannot be estimated.
 \end{enumerate}
\item {\bf Spacecraft separation setup:} The 3D solar spectroscopy methodology developed here can be applied efficiently over a wide range $10\deg$ - $100\deg$ of separation angles between Solar Orbiter SPICE and Hinode/EIS or IRIS for practically the entire duration of the Solar Orbiter mission, subject to observation of the same area in the Sun. The most favourable for spacecraft separation is expected to be $90\deg$.

\end{enumerate}

The first attempt of Doppler maps stereoscopic observations in coronal lines was conducted by the consortium  during Solar Orbiter cruise phase on 21-28 April 2021, as part of a quadrature observation campaign. Test data for vector velocity reconstruction will be available for the community shortly after calibration. Velocity vectors in plasma flows propagating at small  angles to the observation plane (the epipolar plane, where the LOS Doppler velocities are located) can be reconstructed  with higher accuracy,  but the triangulation error may increase.  The high spectral and spatial resolution of the instruments also contributes to an increase in the accuracy of the 3D velocity reconstruction, i.e., Solar Orbiter observations in perihelion positions minimise triangulation errors. The applicability of the methodology is determined by the correct solution of two problems, stereoscopic tie-pointing and triangulation of features, and application of analytical geometry in space.  All these facts must be taken into account when planning stereoscopic observations using Solar Orbiter, and the detailed error assessment will be provided upon the algorithm test with real stereoscopic Doppler maps data sets.

%% Acknowledgements
%

\begin{acknowledgements}
 The authors acknowledge the anonymous referee and the Editor for providing valuable suggestions and comments. We  are grateful  to the EUI, SPICE  instrument teams and to ESA Solar Orbiter Remote Sensing working group for detailed discussions during the consortium meetings, to Hardi Peter and Bart de Pontieu for  discussions about the wavelengths and stereoscopic observations of Doppler maps. OP is grateful to Conrad Schwanitz for sharing his data analysis experience, William Thompson for the vector velocity definition velocity  and to Anik de Groof for Solar Orbiter orbits and observations planning. OP acknowledges funding from Karbacher Fonds. KB  acknowledges funding from SNSF. CM acknowledges grants PICT 2016-0221 (ANPCyT) and UBACyT 20020170100611BA. CM is a member of the Carrera del Investigador Cient\'\i fico of the Consejo Nacional de Investigaciones Cient\'\i ficas y T\'ecnicas (CONICET). The authors thank the Belgian Federal Science Policy Office (BELSPO) for the provision of financial support in the framework of the PRODEX Programme of the European Space Agency (ESA) under contract numbers PEA 4000112292 and 4000117262.
\end{acknowledgements}

   \bibliographystyle{aa} % style aa.bst
 %  \bibliography{bibliography5.bib} % your references Yourfile.bib


%%%%%%%%%%%%%%%%%%%%%%%%%%%%%%%%%%%%%%%%%%%%%
\end{document}